\documentclass[aps,prl,twocolumn,showpacs,superscriptaddress,groupedaddress,showpacs]{revtex4}  

\usepackage{graphicx}  
\usepackage{dcolumn}   
\usepackage{pbox}
\usepackage{bm}        
\usepackage{amssymb}   
\usepackage{amsmath}

\begin{document}

\renewcommand{\arraystretch}{1.5}
\mathchardef\mhyphen="2D

\title{Sympathetic Ground State Cooling and Time-Dilation Shifts in an $^{27}\!\text{Al}^+$ Optical Clock}

\author{{J.-S. Chen}}
\affiliation{Time and Frequency Division, National Institute of Standards and Technology, Boulder, Colorado 80305, USA}
\affiliation{Department of Physics, University of Colorado, Boulder, Colorado 80309, USA}

\author{{S. M. Brewer}}
\affiliation{Time and Frequency Division, National Institute of Standards and Technology, Boulder, Colorado 80305, USA}

\author{{C. W. Chou}}
\affiliation{Time and Frequency Division, National Institute of Standards and Technology, Boulder, Colorado 80305, USA}

\author{{D. J. Wineland}}
\affiliation{Time and Frequency Division, National Institute of Standards and Technology, Boulder, Colorado 80305, USA}
\affiliation{Department of Physics, University of Colorado, Boulder, Colorado 80309, USA}

\author{{D. R. Leibrandt}}
\email[Electronic address: ]{david.leibrandt@nist.gov}
\affiliation{Time and Frequency Division, National Institute of Standards and Technology, Boulder, Colorado 80305, USA}
\affiliation{Department of Physics, University of Colorado, Boulder, Colorado 80309, USA}

\author{{D. B. Hume}}
\email[Electronic address: ]{david.hume@nist.gov}
\affiliation{Time and Frequency Division, National Institute of Standards and Technology, Boulder, Colorado 80305, USA}

\date{\today}

\begin{abstract}
We report on Raman sideband cooling of ${^{25}\text{Mg}^+}$ to sympathetically cool the secular modes of motion in a ${^{25}\text{Mg}^+\,\mhyphen\,^{27}\!\text{Al}^+}$ two-ion pair to near the three-dimensional (3D) ground state.
The evolution of the Fock-state distribution during the cooling process is studied using a rate-equation simulation, and various heating sources that limit the efficiency of 3D sideband cooling in our system are discussed.
We characterize the residual energy and heating rates of all of the secular modes of motion and  estimate a secular motion time-dilation shift of ${-(1.9 \pm 0.1)\times 10^{-18}}$ for an ${^{27}\text{Al}^+}$ clock at a typical clock probe duration of $150$ ms. This is a 50-fold reduction in the secular motion time-dilation shift uncertainty in comparison with previous ${^{27}\text{Al}^+}$ clocks.
\end{abstract}


\maketitle

Trapped and laser-cooled ions are useful for many applications in quantum information processing and quantum metrology because of their isolation from the ambient environment and mutual Coulomb interaction~\cite{Blatt2008Nature,Rosenband2008Science,Lanyon2011Science,Monroe2013Science,Esslinger2013NJP,Monz2016Science}. The Coulomb interaction establishes normal modes of motion, called secular modes, which enable information exchange and entanglement between ions.
For many operations in quantum information processing and metrology, these secular modes should ideally be prepared in their ground state.
For example, in state-of-the-art trapped-ion optical clocks~\cite{Ludlow2015RMP}, uncertainty in the Doppler shift due to the residual excitation of these modes is a dominant contribution to the total clock uncertainty~\cite{ChouAlAlcomparison,Barwood2014PRA,Huntemann2016PRL}.

All trapped-ion optical clocks to date have been operated with the ions' motion near the Doppler cooling limit~\cite{ChouAlAlcomparison,Barwood2014PRA,Ludlow2015RMP,Huntemann2016PRL}.
For clocks based on quantum-logic spectroscopy of the ${^1S_0 \leftrightarrow\, ^3\!P_0}$ transition in ${^{27}\text{Al}^+}$~\cite{Schmidt2005Science}, the smallest uncertainty has been achieved by performing continuous sympathetic Doppler cooling on the logic ion ($^{25}\text{Mg}^+$) during the clock interrogation~\cite{ChouAlAlcomparison}.
Due to the difficulty of performing accurate temperature measurements of trapped ions near the Doppler cooling limit, the uncertainty of the secular motion temperature was limited to $30\,\%$~\citep{ChouAlAlcomparison}.

To reduce the secular motion time-dilation shift and its uncertainty, sub-Doppler cooling techniques can be employed~\cite{Diedrich1989PRL,Monroe1995PRL, Roos2000PRL,Manfredi2012PRL, Lin2013PRL,Ejtemaee2016SisyphusCooling}.
These schemes have not previously been implemented in ion-based clock experiments due to high ambient motional heating rates, the need for extra cooling laser beams, and the difficulty of characterizing the resulting motional state distribution, which can be non-thermal~\cite{ChouAlAlcomparison,NisbetJones2016APB,Huntemann2016PRL}.
For example, in sideband cooling, a non-thermal distribution can result from zero-crossings in the cooling transition Rabi rate as a function of the Fock state~\cite{Poulsen2012PRA,Wan2015PRA} and from cooling durations insufficient to reach equilibrium.
Although sideband cooling can reach extremely low energies, a small non-thermal component of the distribution may contribute more than $90\,\%$ of the total energy~\cite{SM}, rendering the temperature measurement technique using motional sideband ratios unsuitable for determining the energy~\cite{Diedrich1989PRL,Monroe1995PRL}.

Here we demonstrate sympathetic cooling of a ${^{25}\text{Mg}^+\,\mhyphen\,^{27}\!\text{Al}^+}$ two-ion pair to the 3D motional ground state with high probability.
This is accomplished with Raman sideband cooling, selected for its low cooling limit and experimental convenience.
The ion trap used here has motional heating rates less than $12$ $\text{quanta}/\text{s}$~\cite{SM}, two orders of magnitude lower than those achieved in previous ${^{27}\text{Al}^+}$ clocks~\cite{ChouAlAlcomparison}.
The evolution of the Fock state distribution during the cooling process is investigated by using rate-equation simulations which is verified by comparison with experimental measurements.
The cooling pulse sequence is optimized according to the model. We discuss the various heating mechanisms and the secular motion time-dilation shift uncertainty in this system.

\begin{figure}
\includegraphics{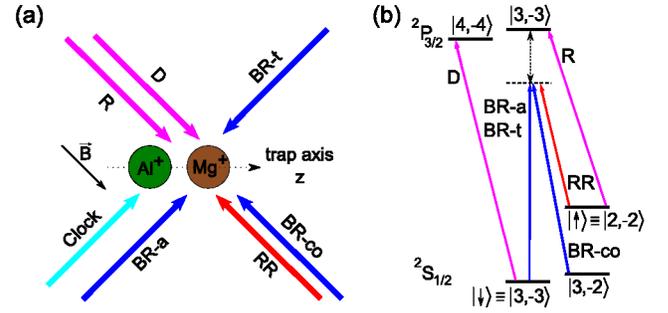}
\caption{\label{wide_f_scan}
(a) Geometry of laser beams. (b) ${^{25}\text{Mg}^+}$ energy level diagram (not to scale). Quantum states ${|F, m_F\rangle}$ are labeled by the total angular momentum $F$ and the projection along the quantization direction $m_F$. The Raman laser detuning from the transition ${|\!\!\downarrow\!\!(\uparrow)\rangle\leftrightarrow|^2P_{3/2},\,F = 3,\,m_F=-3\rangle}$ is approximately 50 GHz and the magnetic field is about 1 G. D, Doppler cooling; R, repumping; RR, red Raman; BR-\textit{a}, axial blue Raman; BR-\textit{t}, transverse blue Raman; BR-co, co-propagating blue Raman. (Clock) $^{27}\text{Al}^+$ laser beam driving the ${^1S_{0}\leftrightarrow^3\!P_0}$ clock transition.
}
\end{figure}

One ${^{25}\text{Mg}^+}$ and one ${^{27}\text{Al}^+}$ are trapped simultaneously in a linear Paul trap to form a pair aligned along the trap (z) axis. The energy levels of ${^{25}\text{Mg}^+}$ and laser beam geometry are shown in Fig.~\ref{wide_f_scan}.
Mutually orthogonal directions x and y are transverse to z.
Along each axis there are two motional modes, the ``common'' (COM) mode where mode vectors of the two ions are in the same direction and the ``stretch'' (STR) mode where they are opposed.
Our trap is operated with axial mode frequencies of $2.7 \,\text{MHz}$ (COM) and $4.8 \,\text{MHz}$ (STR) and transverse mode frequencies in the range of $5$ to $7.5 \,\text{MHz}$~\cite{SM}.
We begin each experiment with $2\,\text{ms}$ of Doppler cooling on the ${^{25}\text{Mg}^+}$ ${|\!\!\downarrow\rangle} \equiv {|^2S_{1/2},\,F=3,\,m_F=-3\rangle} \leftrightarrow {|^2P_{3/2},\,F=4,\,m_F=-4\rangle}$ cycling transition to cool all motional modes close to the Doppler limit and prepare $^{25}\text{Mg}^+$ in ${|\!\!\downarrow\rangle}$.
The subsequent sideband cooling and motional state diagnosis are implemented with stimulated Raman transitions between the ${|\!\!\downarrow\rangle}$ and ${|\!\!\uparrow\rangle} \equiv {|^2S_{1/2},\,F=2,\,m_F=-2\rangle}$ states. 
For efficient cooling of all modes, we employ two sets of Raman beams (\mbox{Fig.~\ref{wide_f_scan} (a)}).
The axial (BR-a) and transverse (BR-t) blue Raman beams in combination with the red Raman beam (RR) generate the differential wavevectors to cool the axial and transverse modes of motion respectively.
Raman laser beams are transmitted through optical fibers for mode filtering and reduction of beam-pointing fluctuations~\cite{Clolombe2014OE}. 
Raman ``carrier'' pulses denote transitions with no change in motional quanta, while red-sideband (RSB) and blue-sideband (BSB) pulses correspond to ${|\!\!\downarrow, n\rangle\!\rightarrow\!|\!\!\uparrow, n\!-\!\Delta n\rangle}$ and ${|\!\!\downarrow, n\rangle\!\rightarrow\!|\!\!\uparrow, n\!+\!\Delta n\rangle}$ transitions respectively, where $n$ is the initial Fock state and $\Delta n$, the order of the sideband transition, represents the number of quanta changed in a single pulse.
After a RSB cooling transition, $^{25}\text{Mg}^+$ is in ${|\!\!\uparrow\rangle}$ which is then optically pumped to ${|\!\!\downarrow\rangle}$ by applying a pulse sequence with three pulses that drive the ${|\!\!\uparrow\rangle}$ to ${|^2P_{3/2},\,F=3,\,m_F=-3\rangle}$ transition, between which are inserted two Raman carrier pulses that drive the ${|^2S_{1/2},\,F=3,\,m_F=-2\rangle}$ to ${|\!\!\uparrow\rangle}$ transition by applying RR and BR-co simultaneously.
The $^{25}\text{Mg}^+$ atomic state is determined at the end of each experiment by collecting photons scattered from the cycling transition and converting the photon count histogram to the transition probability~\cite{SM}.

In general, when the Lamb-Dicke limit is not rigorously satisfied, resolved sideband cooling becomes complicated due to the presence of Fock-state population ``traps'' where the Rabi rate of the RSB transition approaches zero~\cite{Poulsen2012PRA,Wan2015PRA}.
For example, the 1\textsuperscript{st} order RSB Rabi rate for the z-COM mode (Lamb-Dicke parameter $\eta = 0.18$) nearly vanishes for $n = 44$.
Numerical simulations indicate that the 2\textsuperscript{nd} order RSB pulses are significantly more efficient at reducing populations in high Fock states, depending on the Lamb-Dicke parameter and the time for repumping during each cooling cycle.
Multiple 2\textsuperscript{nd} order RSB cooling pulses are applied before the 1\textsuperscript{st} order RSB pulses for the two axial modes in our experiment~\cite{SM}.
After sideband cooling, all RSB signals nearly vanish, indicating the two-ion pair is close to the 3D motional ground state~\cite{SM}.

\begin{figure}
\includegraphics{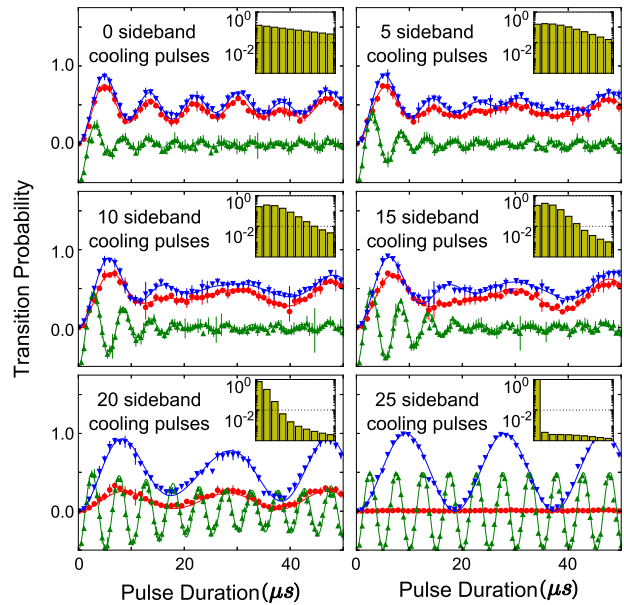}
\caption{\label{pbp_cooling}
Rabi oscillations of the first RSB (red, circle), the first BSB (blue, down triangle) and carrier (green, up triangle) Raman transitions of the axial mode for a single $^{25}\text{Mg}^+$ after different numbers of pulses in the sideband cooling sequence.
Carrier Rabi-oscillation curves are shifted by $-0.5$ for visibility. The cooling sequence consists of 17 second order RSB pulses followed by eight first order RSB pulses to prepare the ion near the motional ground state. Solid lines are given by the rate-equation simulation without any free parameters~\cite{SM}.
The simulated population evolution of the first ten Fock states during the cooling process are shown in the insets.
We include $150$ Fock states in the simulation.
}
\end{figure}

Residual motional excitation can be determined by measuring the ratio of the BSB and RSB transition probabilities~\cite{Diedrich1989PRL, Monroe1995PRL}.
However, this method is accurate only in thermal equilibrium, which is generally not true after sideband cooling~\cite{Barrett2003PRA}.
To overcome this problem, we develop a rate equation model to simulate the sideband cooling process.
We model the cooling dynamics using the coherent RSB transition probability~\cite{WinelandBible}
\begin{equation}
P_{\uparrow,n}(t) = \frac{1}{2}[ 1- e^{-\gamma  t}\cos(2\,\Omega_{n,n+\Delta n}\,t) ]P_{\downarrow,n+\Delta n}(0) \text{,}
\end{equation} 
where $\Delta n$ is the order of the RSB pulse, $\Omega_{n,n+\Delta n}$ is the Rabi rate for the transition ${|\!\!\downarrow,\,n+\Delta n\rangle \rightarrow |\!\!\uparrow,\,n\rangle}$, $\gamma$ is the decoherence rate, and $t$ is the pulse duration.
Following Ref.~\cite{Turchette2000PRA}, we treat the heating as an interaction with a thermal reservoir.
We validate our model by comparing it with experimental data at intermediate times during the cooling sequence. Separate calibrations for the Rabi rates, decoherence rates, and Lamb-Dicke parameters, in combination with the Fock state distributions from our simulation, are used to produce Rabi-oscillation curves of the secular motion along the z direction for a single $^{25}\text{Mg}^+$ with no free parameters.
As shown in Fig.~\ref{pbp_cooling}, Rabi oscillations from our simulation yield good agreement with experimental data throughout the sideband cooling process.
Simulations also reveal the non-thermal distribution after sideband cooling~\cite{SM}, and indicate the energy of ions after sideband cooling may be underestimated by orders of magnitude in certain situations if it is determined using the BSB-RSB ratio method.

During the sideband cooling process, ions experience recoil from both Raman and repumping pulses. The recoil heating due to Raman beams arises from transitions through the ${|^2P_{3/2}\rangle}$ excited states~\cite{Ozeri2007PRA}.
The heating rate due to Raman scattering is estimated to be ${<10^{-4}\, \text{quanta}/\mu s}$ for each motional mode~\cite{SM}, both theoretically and via calibration experiments on a single $^{25}\text{Mg}^+$ for a given Raman beam intensity~\cite{Wayne1982PRA}.
The Rayleigh scattering rate due to the Raman beams is estimated to be $50\,\%$ of the Raman scattering rate~\cite{Ozeri2007PRA}.
The resulting heating is independent of the frequency difference between Raman beams and all motional modes will heat each time the Raman beams are applied to the ions. For this reason, while the heating rates are small this mechanism contributes significantly to the ions' final energy.

The recoil heating in the repumping sequence is measured by preparing ions in the motional ground state and then applying a carrier pulse followed by a repumping sequence multiple times.
This heating scales with $\eta^2$ and is about $0.027$ quanta per cycle in the z-COM mode. Although this value is relatively large, the effective heating due to repumping needs to be multiplied by the probability of population not in the ground state, which becomes negligible at the end of the sideband cooling process.

\begin{figure}
\includegraphics{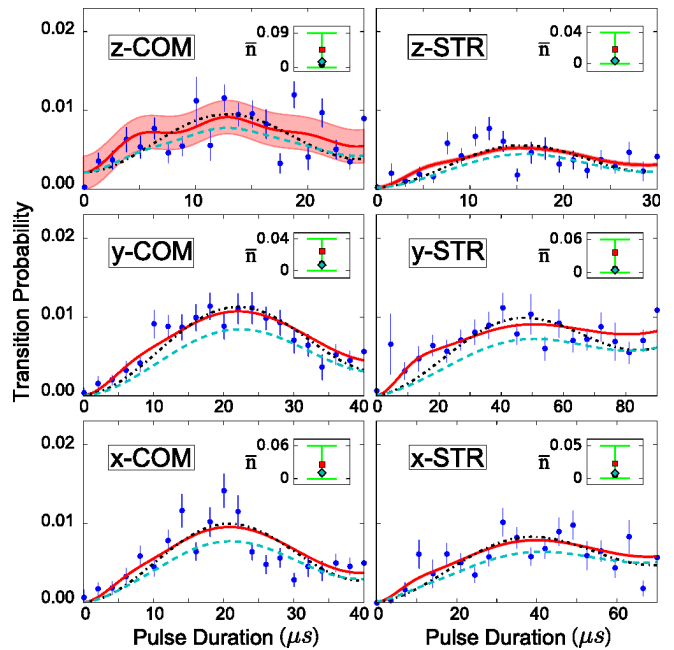}
\caption{\label{rsb_flops}
Red-sideband Rabi oscillations on the ${|\downarrow\rangle\rightarrow|\uparrow\rangle}$ transition of $^{25}\text{Mg}^+$ for the six ${^{25}\!\text{Mg}^+\,\mhyphen\,^{27}\!\text{Al}^+}$ motional modes after Raman sideband cooling. 
The blue data points are the average of about $50\,000$ experiments and the error bars are the standard deviation of the mean.
Solid line, double-thermal distribution fit; dotted line, single-thermal distribution fit; dashed line: rate-equation simulation. The red shaded regions represent the range of the off-resonant carrier transitions of the double-thermal distribution fit, which is significant for the \textit{z}-COM mode, but not the other motional modes~\cite{SM}.
The insets represent the average occupation numbers from fits and the simulation.
Diamond, rate-equation simulation; circle: single thermal distribution fit; square, double thermal distribution fit. Circles overlay diamonds to within less than the size of the symbol. The green error bars represent the experimental uncertainties of average occupation numbers after sideband cooling.
The upper bound of energy is given by the 95\% confidence interval of the double-thermal distribution fit. 
}
\end{figure}

Effects of off-resonant coupling to the carrier transition during RSB pulses are discussed in~\cite{Monroe1995PRL}.
For an ion in the motional ground state, the probability of motional excitation due to this mechanism scales as $1/\omega^2$, where $\omega$ is the motional frequency.
Another important source of heating results from the combination of spontaneous emission due to Raman beams and the RSB pulses.
After a spontaneous decay to ${|\!\!\uparrow\rangle}$, the RSB pulse will add one motional quantum to the mode addressed.
The probability of motional excitation due to this mechanism depends on the RSB pulse duration and scales as $1/\eta$.
This mechanism is only present when Raman transitions are used for sideband cooling; optical transitions are immune to this effect.
Heating from electric field noise is measured as $\dot{\bar{n}} \approx 10\,\text{quanta}/\text{s}$ and $1\,\text{quanta}/\text{s}$ for the COM and STR modes respectively~\cite{SM}.
The heating due to off-resonant coupling to the BSB transitions is not considered in our simulation because of its relatively small transition rate which scales as $\eta^2/\omega^2$.

To experimentally constrain the motional energy after Raman sideband cooling, we focus on Rabi oscillations of RSB transitions to investigate the population not in the ground state.
Motivated by the simulation results, we fit our experimental data to a linear combination of two thermal distributions~\cite{SM}.
Simulation results, fits, and experimental data are plotted in Fig.~\ref{rsb_flops}.
The mean occupation number, $\bar{n}$, extracted from different methods and the energy uncertainties after sideband cooling are shown in insets.
The upper bound of this uncertainty is given by the $95\,\%$ confidence interval of the fit to the double thermal distribution, which gives the highest energy compared to our simulation and the estimate from single thermal distribution fit. The lower bound is set to zero. We do not claim the double thermal distribution is a complete description of the experimental data, but we think it provides a conservative upper limit on the time-dilation shift due to the secular motion.

\begin{figure}
\includegraphics{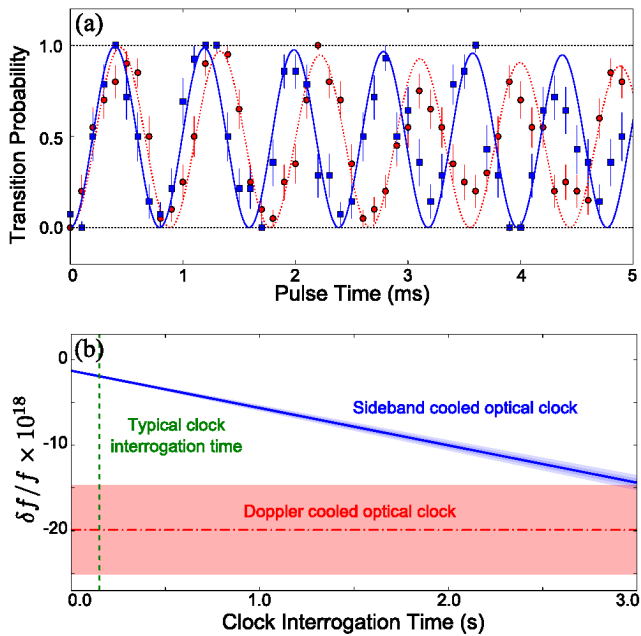}
\caption{\label{TD_cock_flops}
(a) Rabi oscillation of the $^{27}\text{Al}^+\,{^1S_0 \leftrightarrow ^3\!P_0}$ clock transition.
Red circles, continuous sympathetic Doppler cooling, ${\Omega_\text{Doppler} = 2\pi \times 1.126(3) \,\text{kHz}}$; blue squares, 3D sideband cooling, ${\Omega_\text{sc} = 2\pi \times 1.258(6) \,\text{kHz}}$.
Lines are from fitting to an exponential decaying sinusoidal function.
(b) Comparison of the secular motion time-dilation shifts in two different clock operating conditions.
The solid line and the dash-dotted line represent the fractional frequency shifts, while the shaded regions are uncertainties. 
For a typical clock interrogation time of $150$ ms, the secular motion time-dilation shift of a sideband cooled optical clock is $-(1.9 \pm 0.1)\times 10^{-18}$. The uncertainty of a Doppler cooled optical clock is the sum of the secular motion time-dilation shift, $-(16.3 \pm 5.0) \time 10^{-18}$, and the ac Stark shift from the Doppler cooling light, $-(3.6 \pm 1.5) \time 10^{-18}$~\cite{ChouAlAlcomparison}.
}
\end{figure}

To further verify that all motional modes of the ${^{25}\text{Mg}^+\,\mhyphen\,^{27}\!\text{Al}^+}$ two-ion pair are cooled close to the zero-point energy, we compare the Rabi oscillation frequencies of the $^{27}\text{Al}^+{\,\,^1S_0 \leftrightarrow ^3\!P_0}$ clock transition in two conditions and show results in Fig.~\ref{TD_cock_flops}(a).
We first operate the clock near the Doppler limit by sympathetic Doppler cooling during the clock interrogation~\citep{ChouAlAlcomparison}. 
Since the damping rate from the Doppler cooling is much larger than the Rabi rate of the clock transition, the entire Fock state distribution is averaged over a thermal distribution during a single clock pulse and the Rabi oscillation maintains coherence.
In a second experiment we operate the clock by sideband-cooling all motional modes before the clock interrogation pulse.
We expect the Rabi rate of the Doppler cooled optical clock ($\Omega_{Doppler}$) will be $11(2)\,\%$ smaller than that of the sideband cooled optical clock ($\Omega_{sc}$) due to the Debye-Waller effect~\cite{WinelandBible},
\begin{equation}
\frac{\Omega_{Doppler}}{\Omega_{sc}} = \prod_p e^{-\eta_p^2\bar{n}_p}\text{,}
\end{equation}    
where $\eta_p$ and $\bar{n}_p$ represent the Lamb-Dicke parameter of $^{27}\!\text{Al}^+$ and the average occupation number at the Doppler limit of motional mode $p$ respectively.
The uncertainty of the prediction is due to the uncertainty in the motional energy estimation near the Doppler limit. Experimentally, we observe a $10.5(4)\,\%$ difference between the two clock operating conditions, which agrees with the theoretical prediction.

After preparing ions in the 3D motional ground state, there is no additional cooling light during the clock interrogation and we can eliminate the associated light shift reported in the previous $^{27}\text{Al}^+$ optical clocks~\cite{ChouAlAlcomparison}.
The average occupation number of the ions in a specific motional mode $p$ during a clock interrogation of duration $t_i$ can be expressed as
\begin{equation}
\langle n_p(t_i) \rangle =\bar{n}_{p,0} + \frac{1}{2} \dot{\bar{n}}_p t_i\text{,}
\end{equation}
where $\bar{n}_{p,0}$ is the energy after sideband cooling given in Fig.~\ref{rsb_flops} and $\dot{\bar{n}}_p$ is the heating rate due to the ambient electric field noise.
Angle brackets denote an average over the clock interrogation time.
A comparison of the secular motion time-dilation shift using sideband cooling here with continuous Doppler cooling from Ref.~\cite{ChouAlAlcomparison} is shown in Fig.~\ref{TD_cock_flops}(b).
The heating rates of motional modes are the average of measurements spanning several weeks~\cite{SM}. Using the $2\sigma$ uncertainty of our heating rate data to estimate the secular motion time-dilation shift uncertainty, we expect ${\delta\!f/f = -(1.9\,\pm\,0.1)\times 10^{-18}}$ for a $150$ ms clock interrogation time~\cite{SM}.
The overall frequency shift is dominated by the zero-point energy, which can be accurately determined, and the uncertainty is dominated by the energy uncertainty after sideband cooling.
This is roughly an order of magnitude reduction in the shift and a factor of $50$ reduction in the uncertainty due to secular motion time-dilation in comparison with $-(16.3\,\pm\,5.0)\times 10^{-18}$ reported in the previous $^{27}\text{Al}^+$ optical clock~\cite{ChouAlAlcomparison}.
Extending the clock interrogation time to 1 s~\cite{Chou2011QCoherence,Kessler2012NP,Hafner2015OL,Hume2016PRA}, the fractional time-dilation shift due to secular motion is estimated to be ${-(5.7 \pm 0.3)\times 10^{-18}}$.

In conclusion, we have sympathetically cooled a ${^{25}\text{Mg}^+\,\mhyphen\,^{27}\!\text{Al}^+}$ two-ion pair to near the 3D motional ground state using stimulated Raman sideband cooling.
A rate-equation simulation has been performed to characterize the secular motion energy of the ions in our ion trap, which agrees with experimental data throughout the sideband cooling process.
The fractional frequency shift of the $^{27}\text{Al}^+$ ${^1S_0 \leftrightarrow ^3\!\!P_0}$ clock transition due to secular motion is estimated based on the characterization of ion motion and heating rate measurements. The shift due to secular motion time-dilation is reduced by an order of magnitude while the uncertainty is reduced by a factor of $50$ in comparison with the previous $^{27}\text{Al}^+$ optical clock.
Our model may benefit other experiments utilizing sideband cooling to design efficient cooling pulse sequences.

We thank T. Rosenband for useful discussions.
We thank Y. Lin and T. R. Tan for careful reading of the manuscript.
This work was supported by the Defense Advanced Research Projects Agency and the Office of Naval Research. S.M.B. was supported by the U.S. Army Research Office through a MURI grant W911NF-11-1-0400.
This paper is a contribution of the U.S. government, and not subject to U.S. copyright.

\section{Supplemental Material}

\subsection{The Trap}
\begin{figure}
\includegraphics{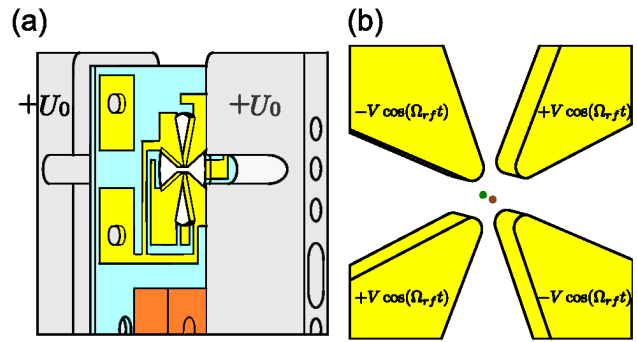}
\caption{\label{trap}
(a) Drawing of the trap electrodes. Grey: endcap electrodes. Light blue: transparent diamond wafer. Yellow: sputtered gold electrodes. Orange: trap support structure. (b) Enlarged view of the four rf electrodes. One $^{25}\text{Mg}^+$ (brown) and one $^{27}\text{Al}^+$ (green) are trapped between the four electrodes at the center of the figure.
}
\end{figure}

Our linear Paul trap is made of a $300\,\mu m$ thick diamond wafer and two titanium endcap electrodes as shown in Fig.~\ref{trap}. The laser-machined, gold-sputtered diamond wafer has four rf electrodes and two micromotion compensation electrodes. The rf electrodes are diagonally connected on the wafer and driven differentially at $76$ MHz to provide the transverse confinement. Two endcap electrodes are mounted about $4$ mm from the ions  and DC biased to provide the axial confinement. The ion-electrode spacing is $250\,\mu m$.

\subsection{Electronic State Detection}
\begin{figure}
\includegraphics{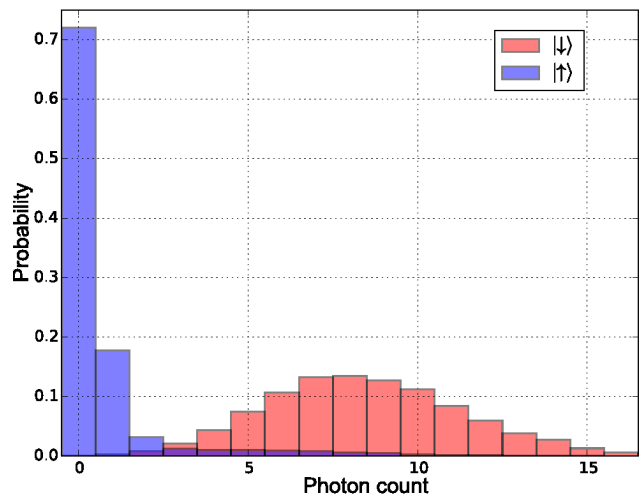}
\caption{\label{ref_hist}
Typical photon histograms scattered from $|\!\!\downarrow\rangle$ and $|\!\!\uparrow\rangle$.
}
\end{figure}

In our experiment, the mean photon counts collected during the detection time of  $250\,\mu s$ from $|\!\!\downarrow\rangle$ and $|\!\!\uparrow\rangle$ are about $8$ and $0.9$, respectively, and the two counts histograms overlap significantly as shown in Fig.~\ref{ref_hist}. The observed histograms in the experiment typically result from the scattered photons from a linear combination of two states and we find that applying a threshold to the photon counts is not sufficient to extract state populations at the $0.1\,\%$ level. To extract the probability in $|\!\!\downarrow\rangle$, we use a maximum-likelihood estimation method to analyze our data. Before experiments start, we need two reference histograms, ${\mathcal{P}(k|\!\!\downarrow)}$ and ${\mathcal{P}(k|\!\!\uparrow)}$, corresponding to the probability of observing $k$ photons in $|\!\!\downarrow\rangle$ and $|\!\!\uparrow\rangle$, respectively. We determine reference histograms experimentally by taking fluorescence measurements when the ions are prepared in the states of $|\!\!\uparrow \rangle$ or $|\!\!\downarrow\rangle$. The state $|\!\!\downarrow\rangle$ is prepared by applying a repumping pulse sequence after Doppler cooling while $|\!\!\uparrow\rangle$ is prepared by applying a microwave $\pi$ pulse from $|\!\!\downarrow\rangle$. During the experiment, we record a series of photon counts $\{k_i\}$ from a particular quantum state ${|\Psi\rangle = \sqrt{p}\,|\!\!\downarrow\rangle + \sqrt{(1-p)}\,e^{i\phi}\,|\!\!\uparrow\rangle}$ in $m$ measurements, where $p$ denotes the probability in $|\!\!\downarrow\rangle$ and $\phi$ is the relative phase between $|\!\!\downarrow\rangle$ and $|\!\!\uparrow\rangle$. The errors in preparation of $|\!\!\downarrow\rangle$, $|\!\!\uparrow\rangle$ and the superposition states are negligible and do not contribute significantly to the determination of the kinetic energy. For a binomial system, the likelihood function is defined as,
\begin{equation}
\mathcal{L}(p) = \prod_{i=1}^m \big(\,p\,\mathcal{P}(k_i|\!\!\downarrow)+(1-p)\,\mathcal{P}(k_i|\!\!\uparrow)\,\big).
\end{equation}
By maximizing the likelihood function numerically, we can estimate the probability in $|\!\!\downarrow\rangle$ for any histogram observed in the experiments~\cite{Gaebler2016HFgate}. 

\subsection{Numerical Simulation}
In the simulation, we treat a RSB pulse duration as a single time step and deal with the heating of the population in $|\!\!\downarrow\rangle$  and $|\!\!\uparrow\rangle$ after the RSB pulse as separate processes. When a $\Delta p_n \mhyphen\,th$ order RSB cooling pulse of duration $t$ is applied to mode $p$, the population in $|\!\!\uparrow\rangle$ is given by
\begin{equation}
\displaystyle
P_{\uparrow}(t) = \sum_{n_p=0}^{\infty} P_{\uparrow, n_p}(t)\text{,}
\end{equation}
where $P_{\uparrow,n_p}$ is given by Eq. (1) in the main text.
During the pulse, ions will heat primarily from the off-resonant carrier transition, and the combination of spontaneous decay and the RSB pulse, which are discussed in the main text. 
Since we are primarily interested in the final steady state, to include these two mechanisms in our model, we calculate the motional excitation rate from $|\!\!\downarrow, n=0 \rangle$ by solving the optical Bloch equation and only including quantum states $|\!\!\downarrow, n=0 \rangle$, $|\!\!\downarrow, n=1 \rangle$, and $|\!\!\uparrow, n=0 \rangle$. The heating values are summarized in Table~\ref{heating}.
This assumption will affect the sideband cooling time constant but not the achievable cooling limit because most of the population is in $|\!\!\downarrow, n=0 \rangle$ and $|\!\!\downarrow, n=1 \rangle$ at the end of the cooling process. 
These two mechanisms mainly affect the mode $p$ while the heating for the rest of the motional modes is through Raman and Rayleigh scattering during the pulse. In addition, all modes will experience recoil heating due to the repumping sequence. We include $150$ Fock states in each of the motional modes in our simulation.
After the repumping sequence, the ion is in $|\!\!\downarrow\rangle$ and we apply additional heating to account for the electric field noise heating, which depends on heating rates and the total time spent on both the RSB pulse and the repumping process. Rabi rates are updated after a RSB pulse based on the Fock state distribution of secular modes to account for the Deybe-Waller effect.

We make following assumptions in the simulation,

\begin{itemize}
\item All the pulses in the cooling and the repumping sequences are perfectly on resonance and applied to the ion for exactly the duration specified.
\item The heating due to spontaneous emission followed by the RSB pulses and the off-resonant coupling are the same for all the Fock states. 
\item The heating due to the scattering of the $|\!\!\downarrow\rangle$ state in a repumping sequence is negligible.
\item The sideband cooling starts from the theoretical Doppler temperature limit.
\item The population in Fock states above ${\text{n}\,=\,150}$ contribute negligibly to the motional energy and can be excluded from the simulation.
\item The $^{25}\text{Mg}^+$ ion stays within the manifold spanned by $|\!\!\uparrow\rangle$, $|\!\!\downarrow\rangle$, and ${|F = 3,m_F = -2\rangle}$ during the cooling process and there is no population leakage to any other states.
\end{itemize}

These assumptions have been verified either numerically or experimentally to ensure they are valid in this experiment. Care should be taken if these assumptions are violated in other experiments.

\subsection{Cooling Pulse Optimization}
Using the model described above, we explored various optimizations of the cooling sequence.  We did not attempt a global optimization of all of the pulse times and frequencies because of the numerous degrees of freedom, which makes it numerically difficult.  Instead we used a pulse-by-pulse optimization, where the total number of pulses is fixed and, for each pulse, the optimizer chooses a sideband (1st-order or 2nd-order) and tries to minimize some figure of merit.  In a global optimization the figure of merit is the total kinetic energy, but in a pulse-by-pulse optimization this is not a good choice because it tends to target the populations in the low-lying Fock states first, which will be be readily cooled by the subsequent pulses.  We tried various figures of merit and found reasonable results by minimizing
\begin{equation}
\displaystyle
\langle n_p\rangle_{opt} = \sum_{n_p=N}^{\infty} P_{n_p} n_p,
\end{equation} 
which is the total motional energy (in the units of $\hbar\,\omega_p$, where $\omega_p$ is the motional frequency of a secular mode $p$) lying above the Fock state $n_p = N-1$, where $N$ is the number of remaining pulses.  The reasoning is that the remaining $N$ pulses can ideally transfer all population in the states from $n_p = N$ to $n_p = 1$ to the ground state by choosing the proper duration of 1st-order sideband pulses.

This optimization was performed for the z-COM mode alone, since that mode is the most problematic in terms of its high Lamb-Dicke parameter and high occupation number at the Doppler limit.  The optimization results suggested a pulse sequence with multiple 2\textsuperscript{nd} order sideband pulses followed by a few 1\textsuperscript{st} order sideband pulses at the end. However, we found that the total kinetic energy after sideband cooling was not very sensitive to the exact number of 2nd-order sideband pulses. Therefore, the cooling sequence that we chose for our experiment was a compromise between the optimization results and other experimental factors such as the cooling efficiency for the other modes and the ease of parameterizing the cooling sequence in our experimental control system.

\begin{figure}
\includegraphics{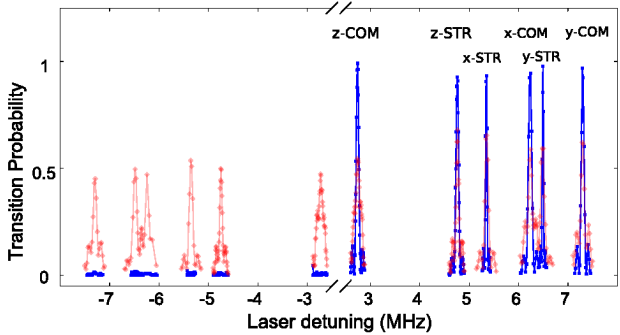}
\caption{\label{raman_spec}
Raman transition spectrum of a $^{25}\text{Mg}^+\,\mhyphen\,^{27}\!\text{Al}^+$ two-ion pair with all six secular modes of motion resolved. 
Red: after Doppler cooling. Blue: after Raman sideband cooling. 
}
\end{figure}

For a single $^{25}\text{Mg}^+$, we apply $17$ 2\textsuperscript{nd} order RSB pulses followed by $8$ 1\textsuperscript{st} order RSB pulses on the axial mode, and $25$ 1\textsuperscript{st} order RSB pulses on two transverse modes. For a $^{25}\text{Mg}^+\,$-$\,^{27}\!\text{Al}^+$ two-ion pair, we apply $15$ 2\textsuperscript{nd} order RSB pulses followed by $25$ 1\textsuperscript{st} order RSB pulsed on the two axial modes, and $40$ 1\textsuperscript{st} order RSB pulses on the transverse modes. The frequency spectra of motional modes before and after the sideband cooling are presented in Fig.~\ref{raman_spec}.

One important conclusion from this process of optimization was that two pulse sequences may result in very different distributions and $\bar{n}$, but yet give very similar results in a typical sideband thermometry experiment.  Using 2\textsuperscript{nd}-order sideband pulses was critical for achieving low kinetic energy in the axial modes with relatively high Lamb-Dicke parameters.  While some more improvement in cooling efficiency may be possible by performing a global optimization, we have concluded that the uncertainty in experimentally verifying the residual energy is much greater than the gains that we might achieve.

\subsection{Ion Motional Energy Estimate}
\begin{figure}
\includegraphics{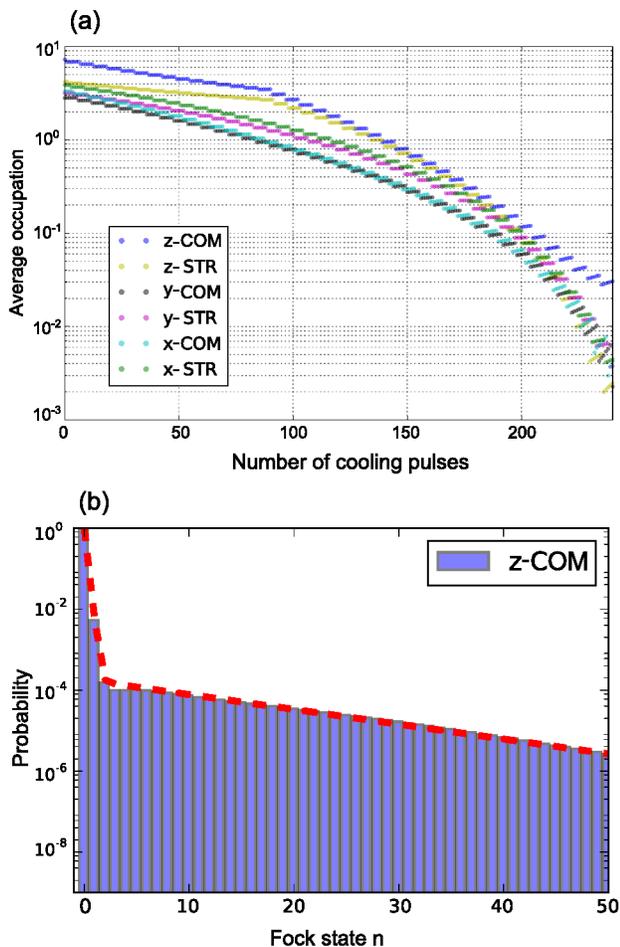}
\caption{\label{Sim_cooling}
(a) Evolution of the mean occupation of the six motional modes from the simulation versus the number of cooling pulses applied. (b) The simulated Fock state distribution in z-COM mode after sideband cooling. The red line represents a fit to a double-thermal distribution with mean occupation numbers $0.004$ and $11$, and weights $0.99$ and $0.01$.
}
\end{figure}

In Fig.~\ref{Sim_cooling} (a), we plot the evolution of the mean Fock state occupation in all six motional modes during the sideband cooling process against the number of cooling pulses applied. For the two axial modes, the different cooling rate occurring around $100$ pulses in the figure corresponds to the change from the 2\textsuperscript{nd} order sideband cooling to the 1\textsuperscript{st} order sideband cooling.
As an example, we show the simulated Fock state distribution of the z-COM mode after sideband cooling in Fig.~\ref{Sim_cooling} (b). Besides the thermal-like distribution for the Fock states $n<3$, our simulation shows a plateau probability distribution for Fock states $ n \gtrsim 5$. Although the sum of the probability in the plateau distribution is too small to be observed in the experiment, our simulation indicates it usually contributes more than $90\,\%$ of the total kinetic energy. In the z-COM mode, about $95\,\%$ of the total energy is in the plateau distribution in the higher Fock states. 

To include the energy contribution from this ``hidden'' population, we fit the simulated Fock state distribution to the linear combination of two thermal distributions to extract the mean occupation number $\bar{n}_h$ characterizing the energy of the plateau distribution. Given $\bar{n}_h$ as a fixed parameter, we then fit the RSB Rabi oscillation data shown in Fig. 3 in the main text to the double thermal distribution. Therefore the probability in the specific Fock state $P(n)$ is expressed as
\begin{equation}
P(n) = \alpha\,P_{th}(n|\bar{n}_l) + (1-\alpha)\,P_{th}(n|\bar{n}_h)\text{,}
\end{equation}
where $\bar{n}_h$ is derived from the simulation to characterize the energy in the higher Fock states and $\bar{n}_l$ characterizes the energy in the lower Fock states. The probability $P_{th}(n|\bar{n})$ represents the population in the Fock state $|n\rangle$ of the thermal distribution with the average occupation $\bar{n}$ and $\alpha$ denotes the weight between the two thermal distributions.

We also consider the effect of the off-resonant carrier transitions in our analysis. The off-resonant transitions cause a small amplitude ($<0.5\%$), high frequency ($\gtrsim\text{MHz}$) oscillation in addition to the residual RSB Rabi oscillation. Because both the residual RSB and off-resonant carrier transitions have small amplitudes and very different frequencies, the transition probabilities at any given pulse duration can be approximated by the sum of these two effects. However we do not have the timing resolution to resolve the off-resonant transitions. To include this effect in our energy estimate, we first record the laser power as a function of time for the experimental pulses and numerically calculate the off-resonant carrier transition amplitudes. Then we assume the fast oscillations due to off-resonant transitions cause an offset towards higher transition probability and scatter of the experimental data. Therefore, we add this additional offset to our fitting model and represent the scatter as the red region in Fig. 3 in the main text. This effect is significant for the data of the z-COM RSB transition but not the other motional modes. This is due to the adiabaticity arising from the relatively long turn-on and turn-off time of the laser pulses ($\sim 100$ ns) in comparison with the oscillation frequencies.       

The double thermal fit result is plotted in Fig.~\ref{Sim_cooling} (b) compared with the simulated population. The least-squared fit to the Rabi oscillation curves using the double thermal distribution model gives a factor of two to three smaller in $\chi^2$ compared to the single thermal distribution model. The $95\,\%$ confidence interval of the fit is used as the estimate of the energy upper bound and the results are shown in Fig. 3 in the main text. A bootstrapping method by re-sampling residuals is also performed to verify the energy upper bound and the results are consistent with that from the least-squares fit~\cite{Diciccio1996SS}. The parameters of all secular modes and the estimates of the occupation number given by different methods are summarized in Table~\ref{mode}. The heating rates due to different mechanisms discussed in the paper are shown in Table~\ref{heating}. We also list the heating contribution due to different mechanisms from the simulation in Table~\ref{heat_contribution}.

\subsection{Heating Rate Measurement}
\begin{figure}
\includegraphics{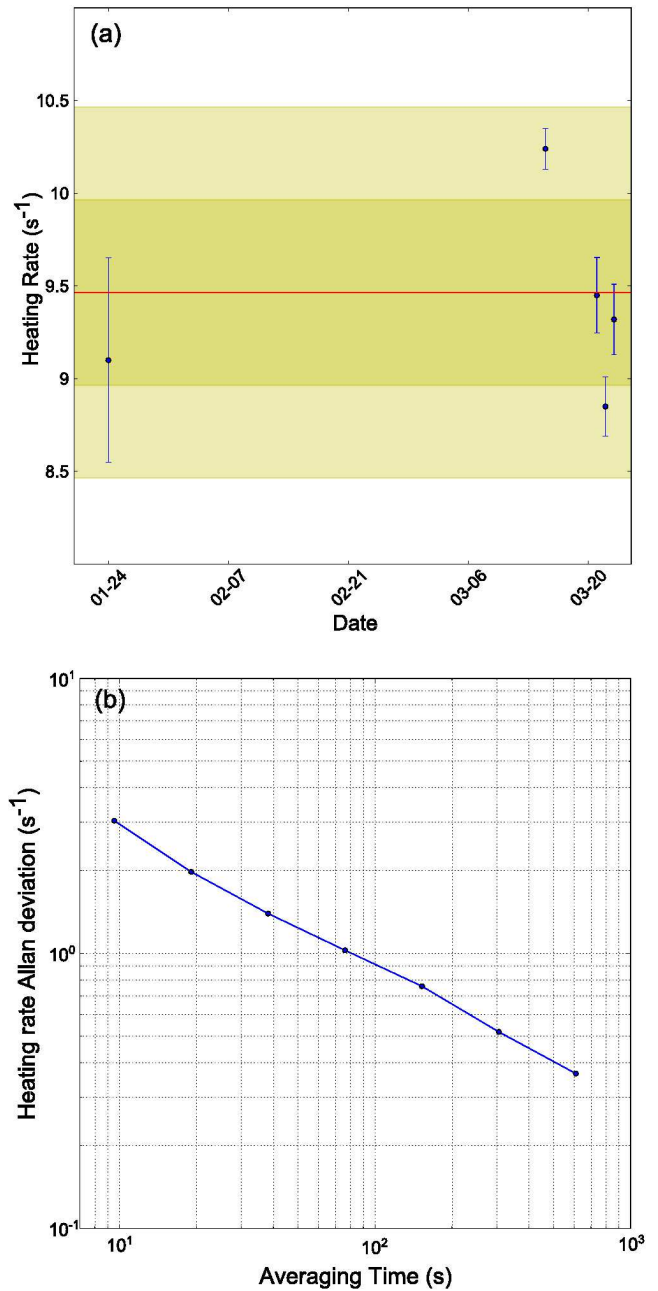}
\caption{\label{HR_stability}
The heating rate of the y-COM mode of motion for a $^{25}\text{Mg}^+\,\mhyphen\,^{27}\!\text{Al}^+$ two-ion pair. (a) The heating rate measurement consisted of five experiments spanning several weeks. Error bars represent the statistical error. The red line and the yellow region are the mean value, $1\sigma$ uncertainty, and $2 \sigma$ uncertainty. (b) Allan deviation of the heating rate calculated from a continuous measurement for about an hour. 
}
\end{figure}

To measure the heating rate of the motion due to the ambient electric field noise, we first sideband-cool the $^{25}\text{Mg}^+\,\mhyphen\,^{27}\!\text{Al}^+$ two-ion pair close to the motional ground state. Depending on the value of the heating rate, ions then experience $10$ to $30$ ms of ``dark time'' without any cooling, such that the ions' motion can  heat due to the ambient electric field noise. After the dark time, we measure the ratio of the RSB and the BSB transition probability to estimate the final occupation number of the ion in a specific motional mode and hence the heating rate. 
This sideband ratio method works well for determining heating rates but not for characterizing ions' energy after sideband cooling because the heating rapidly drives the Fock state distribution towards a thermal distribution~\cite{James1998PRL, Dodonov2000JOB}. As a check of the constancy of the heating rate, repeated measurements of the heating rate of a specific secular mode were made for about an hour. These measurements were then repeated several times over several weeks to verify the long-term stability of the heating. 
The average and the standard error of heating rates are used to estimate the secular motion time-dilation shift. The measurements for one of six motional modes is shown in Fig.~\ref{HR_stability} and the values for all six modes are summarized in  Table~\ref{heating}.

\subsection{Shift and Uncertainty Estimate}
For ions confined in a linear Paul trap, we must account for the intrinsic micromotion in the transverse secular modes of motion~\cite{Wineland1987PRA}. The contribution of the intrinsic micromotion energy  (but not the excess mircomotion energy since it only depends on the trap imperfections) has been included when calculating the time-dilation shift in this paper.
To determine the upper bound on the fractional time-dilation shift for a specific mode $p$ at a given clock interrogation pulse duration $t_i$, we use $(\dot{\bar{n}}_p+2\sigma)$ values from Table~\ref{heating} times $t_i$ added with the upper bound of $\bar{n}_{p,0}$ from Table~\ref{mode}. 
For the lower bound, we take the mode to be initially in its ground state and use $(\dot{\bar{n}}_p-2\sigma)$ value from Table~\ref{heating} times $t_i$ for the contribution from heating.
The bounds of the time-dilation shift of a specific mode $p$ are given as
\begin{align}
\begin{split}
\Big(\frac{\delta f}{f}\Big)^{upper(lower)}_p &= \Big(\frac{\delta\nu _p}{f}\Big) \Big\langle n_p(t_i)^{upper(lower)}\Big\rangle\\
&= \Big(\frac{\delta\nu _p}{f}\Big)\Big( \bar{n}^{upper(lower)}_{p,0}+\frac{1}{2}(\dot{\bar{n}}_p\pm 2\sigma)t_i\Big)\text{,}
\end{split}
\end{align}
where $\delta \nu_{p}$ denotes the secular motion time-dilation shift per quantum and $\langle n_p(t_i)\rangle$ is the calculated average occupation number of the ion secular motion using Eq. (3) in the main text.
Finally, we conservatively assume the shifts in six different secular modes are correlated, which should be the worst case, and use the average of the upper bound and lower bound as the estimate for the time-dilation shift, 
\begin{align}\label{}
&\Big(\frac{\delta f}{f}\Big)^{upper(lower)}_{secular} = \sum _p{\Big(\frac{\delta f}{f}\Big)^{upper(lower)}_p}\text{,}\\
&\Big(\frac{\delta f}{f}\Big)_{secular} = \frac{ (\frac{\delta f}{f})^{upper}_{secular}+(\frac{\delta f}{f})^{lower}_{secular}}{2}\text{,}\\
&\text{uncertainty}= \frac{ (\frac{\delta f}{f})^{upper}_{secular}-(\frac{\delta f}{f})^{lower}_{secular}}{2}\text{.}
\end{align}
The time dilation shift uncertainty due to secular motion in terms of the clock interrogation time is represented by the blue region in Fig. 4(b) in the main text.

\begin{table*}
\caption{\label{mode}
Parameters of secular motion in a $^{25}\text{Mg}^+$-$^{27}\!\text{Al}^+$ two-ion pair in the experiment.
}
\begin{ruledtabular}
\begin{tabular}{ccccccc}
 & x-COM & x-STR & y-COM & y-STR & z-COM & z-STR \\ 
\hline
Frequency (MHz) & 6.2 & 5.4 & 7.2 & 6.5 & 2.7 & 4.8 \\

Lamb-Dicke parameter ($^{25}\text{Mg}^+$)\footnotemark[1] & 0.12 & 0.06 & 0.11 & 0.05 & 0.18 & 0.16 \\

Time-dilation shift per quantum ($10^{-18}$)\footnotemark[2] & $-0.21$ & $-0.85$ & $-0.18$ & $-0.86$ & $-0.12$ & $-0.18$ \\

Occupation number from simulation (quanta) & $0.009\,9$ & $0.009\,0$ & $0.007\,8$ & $0.004\,9$ & $0.015$ & $0.003\,5$ \\

Occupation number from single-thermal distribution fit (quanta) & $0.008\,7$ & $0.005\,9$ & $0.009\,9$ & $0.006\,9$ & $0.006\,6$ & $0.004\,4$ \\

Occupation number from double-thermal distribution fit (quanta) & $0.024$ & $0.024$ & $0.027$ & $0.037$ & $0.044$ & $0.018$ \\

Occupation number bounds (quantum)\footnotemark[3] & $0$ - $0.06$ & $0$ - $0.05$ & $0$ - $0.04$ & $0$ - $0.06$ & $0$ - $0.09$ & $0$ - $0.04$\\

\end{tabular}
\end{ruledtabular}
\footnotetext[1]{calculated values for Raman cooling beams and consistent with experimental values.}
\footnotetext[2]{including the shift due to the intrinsic micromotion in the transverse directions.}
\footnotetext[3]{$95\%$ confidence interval.}
\end{table*}

\begin{table*}
\caption{\label{heating}
Summary of heating rates due to different mechanisms considered in the simulation.
}
\begin{ruledtabular}
\begin{tabular}{ccccccc}
 & x-COM & x-STR & y-COM & y-STR & z-COM & z-STR \\ 
\hline
Repumping sequence (quanta/cycle)\footnotemark[1] & 0.015 & 0.004 & 0.012 & 0.002 & 0.027 & 0.017 \\

Raman scattering (quanta/$\mu s$)\footnotemark[2] & $4.6 \times 10^{-5}$ & $1.2 \times 10^{-5}$ & $4.1 \times 10^{-5}$ & $8.7 \times 10^{-6}$ & $5.8 \times <10^{-5}$ & $3.9 \times 10^{-5}$ \\

Rayleigh scattering (quanta/$\mu s$)\footnotemark[2] & $2.3 \times 10^{-5}$ & $6.0 \times 10^{-6}$ & $2.1 \times 10^{-5}$ & $4.4 \times 10^{-6}$ & $2.9 \times <10^{-5}$ & $2.0 \times 10^{-5}$ \\

Off-resonant coupling (quanta/pulse)\footnotemark[3] & $3.9 \times 10^{-4}$ & $3.6 \times 10^{-4}$ & $2.1 \times 10^{-4}$ & $3.3 \times 10^{-4}$ & $1.6 \times 10^{-3}$ & $6.5 \times 10^{-4}$ \\

Spontaneous decay + RSB (quanta/pulse)\footnotemark[2]\footnotemark[3] & $1.4 \times 10^{-3}$ & $2.6 \times 10^{-3}$ & $1.5 \times 10^{-3}$ & $3.1 \times 10^{-3}$ & $7.5 \times 10^{-4}$ & $9.1 \times 10^{-4}$ \\

Electric field noise (quanta/$s$)\footnotemark[1]\footnotemark[4] & $11.95 \pm 0.64$ & $1.94 \pm 0.08$ & $9.47 \pm 1.00$ & $1.96 \pm 0.18$ & $9.12 \pm 0.32$ & $0.34 \pm 0.02$ \\
\end{tabular}
\end{ruledtabular}
\footnotetext[1]{experimental measurement.}
\footnotetext[2]{estimate using the measured laser power and the beam size.}
\footnotetext[3]{motional excitation from ground state, simulation.}
\footnotetext[4]{$2\sigma$ uncertainty}
\end{table*}

\begin{table*}
\caption{\label{heat_contribution}
The amount of the heating due to different mechanisms in a 3D sideband cooling pulse sequence from the simulation. Unit: quantum.
}
\begin{ruledtabular}
\begin{tabular}{ccccccc}
 & x-COM & x-STR & y-COM & y-STR & z-COM & z-STR \\ 
\hline
Repumping sequence & $5.4 \times 10^{-4}$ & $3.0 \times 10^{-4}$ & $1.7 \times 10^{-4}$ & $3.8 \times 10^{-5}$ & $1.2 \times 10^{-2}$ & $3.2 \times 10^{-4}$ \\

Raman + Rayleigh scattering & $3.0 \times 10^{-3}$ & $2.8 \times 10^{-3}$ & $4.3 \times 10^{-3}$ & $4.2 \times 10^{-3}$ & $ 2.8 \times 10^{-3}$ & $ 1.5 \times 10^{-3}$ \\

Off-resonant coupling & $3.9 \times 10^{-4}$ & $3.6 \times 10^{-4}$ & $2.1 \times 10^{-4}$ & $3.3 \times 10^{-4}$ & $1.6 \times 10^{-3}$ & $6.5 \times 10^{-4}$ \\

Spontaneous decay + RSB & $1.4 \times 10^{-3}$ & $2.6 \times 10^{-3}$ & $1.5 \times 10^{-3}$ & $3.1 \times 10^{-3}$ & $7.7 \times 10^{-4}$ & $9.1 \times 10^{-4}$ \\

Electric field noise & $1.4 \times 10^{-3}$ & $1.8 \times 10^{-4}$ & $2.0 \times 10^{-3}$ & $3.4 \times 10^{-4}$ & $4.3 \times 10^{-4}$ & $1.5 \times 10^{-5}$ \\
\end{tabular}
\end{ruledtabular}
\end{table*}


\begin{thebibliography}{36}%
\makeatletter
\providecommand \@ifxundefined [1]{%
 \@ifx{#1\undefined}
}%
\providecommand \@ifnum [1]{%
 \ifnum #1\expandafter \@firstoftwo
 \else \expandafter \@secondoftwo
 \fi
}%
\providecommand \@ifx [1]{%
 \ifx #1\expandafter \@firstoftwo
 \else \expandafter \@secondoftwo
 \fi
}%
\providecommand \natexlab [1]{#1}%
\providecommand \enquote  [1]{``#1''}%
\providecommand \bibnamefont  [1]{#1}%
\providecommand \bibfnamefont [1]{#1}%
\providecommand \citenamefont [1]{#1}%
\providecommand \href@noop [0]{\@secondoftwo}%
\providecommand \href [0]{\begingroup \@sanitize@url \@href}%
\providecommand \@href[1]{\@@startlink{#1}\@@href}%
\providecommand \@@href[1]{\endgroup#1\@@endlink}%
\providecommand \@sanitize@url [0]{\catcode `\\12\catcode `\$12\catcode
  `\&12\catcode `\#12\catcode `\^12\catcode `\_12\catcode `\%12\relax}%
\providecommand \@@startlink[1]{}%
\providecommand \@@endlink[0]{}%
\providecommand \url  [0]{\begingroup\@sanitize@url \@url }%
\providecommand \@url [1]{\endgroup\@href {#1}{\urlprefix }}%
\providecommand \urlprefix  [0]{URL }%
\providecommand \Eprint [0]{\href }%
\providecommand \doibase [0]{http://dx.doi.org/}%
\providecommand \selectlanguage [0]{\@gobble}%
\providecommand \bibinfo  [0]{\@secondoftwo}%
\providecommand \bibfield  [0]{\@secondoftwo}%
\providecommand \translation [1]{[#1]}%
\providecommand \BibitemOpen [0]{}%
\providecommand \bibitemStop [0]{}%
\providecommand \bibitemNoStop [0]{.\EOS\space}%
\providecommand \EOS [0]{\spacefactor3000\relax}%
\providecommand \BibitemShut  [1]{\csname bibitem#1\endcsname}%
\let\auto@bib@innerbib\@empty
\bibitem [{\citenamefont {Blatt}\ and\ \citenamefont
  {Wineland}(2008)}]{Blatt2008Nature}%
  \BibitemOpen
  \bibfield  {author} {\bibinfo {author} {\bibfnamefont {R.}~\bibnamefont
  {Blatt}}\ and\ \bibinfo {author} {\bibfnamefont {D.~J.}\ \bibnamefont
  {Wineland}},\ }\href@noop {} {\bibfield  {journal} {\bibinfo  {journal}
  {Nature}\ }\textbf {\bibinfo {volume} {453}},\ \bibinfo {pages} {1008}
  (\bibinfo {year} {2008})}\BibitemShut {NoStop}%
\bibitem [{\citenamefont {Rosenband}\ \emph {et~al.}(2008)\citenamefont
  {Rosenband}, \citenamefont {Hume}, \citenamefont {Schmidt}, \citenamefont
  {Chou}, \citenamefont {Brusch}, \citenamefont {Lorini}, \citenamefont
  {Oskay}, \citenamefont {Drullinger}, \citenamefont {Fortier}, \citenamefont
  {Stalnaker}, \citenamefont {Diddams}, \citenamefont {Swann}, \citenamefont
  {Newbury}, \citenamefont {Itano}, \citenamefont {Wineland},\ and\
  \citenamefont {Bergquist}}]{Rosenband2008Science}%
  \BibitemOpen
  \bibfield  {author} {\bibinfo {author} {\bibfnamefont {T.}~\bibnamefont
  {Rosenband}}, \bibinfo {author} {\bibfnamefont {D.~B.}\ \bibnamefont {Hume}},
  \bibinfo {author} {\bibfnamefont {P.~O.}\ \bibnamefont {Schmidt}}, \bibinfo
  {author} {\bibfnamefont {C.~W.}\ \bibnamefont {Chou}}, \bibinfo {author}
  {\bibfnamefont {A.}~\bibnamefont {Brusch}}, \bibinfo {author} {\bibfnamefont
  {L.}~\bibnamefont {Lorini}}, \bibinfo {author} {\bibfnamefont {W.~H.}\
  \bibnamefont {Oskay}}, \bibinfo {author} {\bibfnamefont {R.~E.}\ \bibnamefont
  {Drullinger}}, \bibinfo {author} {\bibfnamefont {T.~M.}\ \bibnamefont
  {Fortier}}, \bibinfo {author} {\bibfnamefont {J.~E.}\ \bibnamefont
  {Stalnaker}}, \bibinfo {author} {\bibfnamefont {S.~A.}\ \bibnamefont
  {Diddams}}, \bibinfo {author} {\bibfnamefont {W.~C.}\ \bibnamefont {Swann}},
  \bibinfo {author} {\bibfnamefont {N.~R.}\ \bibnamefont {Newbury}}, \bibinfo
  {author} {\bibfnamefont {W.~M.}\ \bibnamefont {Itano}}, \bibinfo {author}
  {\bibfnamefont {D.~J.}\ \bibnamefont {Wineland}}, \ and\ \bibinfo {author}
  {\bibfnamefont {J.~C.}\ \bibnamefont {Bergquist}},\ }\href@noop {} {\bibfield
   {journal} {\bibinfo  {journal} {Science}\ }\textbf {\bibinfo {volume}
  {319}},\ \bibinfo {pages} {1808} (\bibinfo {year} {2008})}\BibitemShut
  {NoStop}%
\bibitem [{\citenamefont {Lanyon}\ \emph {et~al.}(2011)\citenamefont {Lanyon},
  \citenamefont {Hempel}, \citenamefont {Nigg}, \citenamefont {M\"uller},
  \citenamefont {Gerritsma}, \citenamefont {Z\"ahringer}, \citenamefont
  {Schindler}, \citenamefont {Barreiro}, \citenamefont {Rambach}, \citenamefont
  {Kirchmair}, \citenamefont {Hennrich}, \citenamefont {Zoller}, \citenamefont
  {Blatt},\ and\ \citenamefont {Roos}}]{Lanyon2011Science}%
  \BibitemOpen
  \bibfield  {author} {\bibinfo {author} {\bibfnamefont {B.}~\bibnamefont
  {Lanyon}}, \bibinfo {author} {\bibfnamefont {C.}~\bibnamefont {Hempel}},
  \bibinfo {author} {\bibfnamefont {D.}~\bibnamefont {Nigg}}, \bibinfo {author}
  {\bibfnamefont {M.}~\bibnamefont {M\"uller}}, \bibinfo {author}
  {\bibfnamefont {R.}~\bibnamefont {Gerritsma}}, \bibinfo {author}
  {\bibfnamefont {F.}~\bibnamefont {Z\"ahringer}}, \bibinfo {author}
  {\bibfnamefont {P.}~\bibnamefont {Schindler}}, \bibinfo {author}
  {\bibfnamefont {J.~T.}\ \bibnamefont {Barreiro}}, \bibinfo {author}
  {\bibfnamefont {M.}~\bibnamefont {Rambach}}, \bibinfo {author} {\bibfnamefont
  {G.}~\bibnamefont {Kirchmair}}, \bibinfo {author} {\bibfnamefont
  {M.}~\bibnamefont {Hennrich}}, \bibinfo {author} {\bibfnamefont
  {P.}~\bibnamefont {Zoller}}, \bibinfo {author} {\bibfnamefont
  {R.}~\bibnamefont {Blatt}}, \ and\ \bibinfo {author} {\bibfnamefont {C.~F.}\
  \bibnamefont {Roos}},\ }\href@noop {} {\bibfield  {journal} {\bibinfo
  {journal} {Science}\ }\textbf {\bibinfo {volume} {334}},\ \bibinfo {pages}
  {57} (\bibinfo {year} {2011})}\BibitemShut {NoStop}%
\bibitem [{\citenamefont {Monroe}\ and\ \citenamefont
  {Kim}(2013)}]{Monroe2013Science}%
  \BibitemOpen
  \bibfield  {author} {\bibinfo {author} {\bibfnamefont {C.}~\bibnamefont
  {Monroe}}\ and\ \bibinfo {author} {\bibfnamefont {J.}~\bibnamefont {Kim}},\
  }\href@noop {} {\bibfield  {journal} {\bibinfo  {journal} {Science}\ }\textbf
  {\bibinfo {volume} {339}},\ \bibinfo {pages} {1164} (\bibinfo {year}
  {2013})}\BibitemShut {NoStop}%
\bibitem [{\citenamefont {Esslinger}\ \emph {et~al.}(2013)\citenamefont
  {Esslinger}, \citenamefont {Schaetz},\ and\ \citenamefont
  {Monroe}}]{Esslinger2013NJP}%
  \BibitemOpen
  \bibfield  {author} {\bibinfo {author} {\bibfnamefont {T.}~\bibnamefont
  {Esslinger}}, \bibinfo {author} {\bibfnamefont {T.}~\bibnamefont {Schaetz}},
  \ and\ \bibinfo {author} {\bibfnamefont {C.}~\bibnamefont {Monroe}},\
  }\href@noop {} {\bibfield  {journal} {\bibinfo  {journal} {New. J. Phys.}\
  }\textbf {\bibinfo {volume} {15}},\ \bibinfo {pages} {085009} (\bibinfo
  {year} {2013})}\BibitemShut {NoStop}%
\bibitem [{\citenamefont {Monz}\ \emph {et~al.}(2016)\citenamefont {Monz},
  \citenamefont {Nigg}, \citenamefont {Martines}, \citenamefont {Brandl},
  \citenamefont {Schindler}, \citenamefont {Rines}, \citenamefont {Wang},
  \citenamefont {Chuang},\ and\ \citenamefont {Blatt}}]{Monz2016Science}%
  \BibitemOpen
  \bibfield  {author} {\bibinfo {author} {\bibfnamefont {T.}~\bibnamefont
  {Monz}}, \bibinfo {author} {\bibfnamefont {D.}~\bibnamefont {Nigg}}, \bibinfo
  {author} {\bibfnamefont {E.~A.}\ \bibnamefont {Martines}}, \bibinfo {author}
  {\bibfnamefont {M.~F.}\ \bibnamefont {Brandl}}, \bibinfo {author}
  {\bibfnamefont {P.}~\bibnamefont {Schindler}}, \bibinfo {author}
  {\bibfnamefont {R.}~\bibnamefont {Rines}}, \bibinfo {author} {\bibfnamefont
  {S.~X.}\ \bibnamefont {Wang}}, \bibinfo {author} {\bibfnamefont {I.~L.}\
  \bibnamefont {Chuang}}, \ and\ \bibinfo {author} {\bibfnamefont
  {R.}~\bibnamefont {Blatt}},\ }\href@noop {} {\bibfield  {journal} {\bibinfo
  {journal} {Science}\ }\textbf {\bibinfo {volume} {351}},\ \bibinfo {pages}
  {1068} (\bibinfo {year} {2016})}\BibitemShut {NoStop}%
\bibitem [{\citenamefont {Ludlow}\ \emph {et~al.}(2015)\citenamefont {Ludlow},
  \citenamefont {Boyd}, \citenamefont {Ye}, \citenamefont {Peik},\ and\
  \citenamefont {Schmidt}}]{Ludlow2015RMP}%
  \BibitemOpen
  \bibfield  {author} {\bibinfo {author} {\bibfnamefont {A.~D.}\ \bibnamefont
  {Ludlow}}, \bibinfo {author} {\bibfnamefont {M.~M.}\ \bibnamefont {Boyd}},
  \bibinfo {author} {\bibfnamefont {J.}~\bibnamefont {Ye}}, \bibinfo {author}
  {\bibfnamefont {E.}~\bibnamefont {Peik}}, \ and\ \bibinfo {author}
  {\bibfnamefont {P.~O.}\ \bibnamefont {Schmidt}},\ }\href@noop {} {\bibfield
  {journal} {\bibinfo  {journal} {Rev. Mod. Phys.}\ }\textbf {\bibinfo {volume}
  {87}},\ \bibinfo {pages} {637} (\bibinfo {year} {2015})}\BibitemShut
  {NoStop}%
\bibitem [{\citenamefont {Chou}\ \emph {et~al.}(2010)\citenamefont {Chou},
  \citenamefont {Hume}, \citenamefont {Koelemeij}, \citenamefont {Wineland},\
  and\ \citenamefont {Rosenband}}]{ChouAlAlcomparison}%
  \BibitemOpen
  \bibfield  {author} {\bibinfo {author} {\bibfnamefont {C.~W.}\ \bibnamefont
  {Chou}}, \bibinfo {author} {\bibfnamefont {D.~B.}\ \bibnamefont {Hume}},
  \bibinfo {author} {\bibfnamefont {J.~C.~J.}\ \bibnamefont {Koelemeij}},
  \bibinfo {author} {\bibfnamefont {D.~J.}\ \bibnamefont {Wineland}}, \ and\
  \bibinfo {author} {\bibfnamefont {T.}~\bibnamefont {Rosenband}},\ }\href@noop
  {} {\bibfield  {journal} {\bibinfo  {journal} {Phys. Rev. Lett.}\ }\textbf
  {\bibinfo {volume} {104}},\ \bibinfo {pages} {070802} (\bibinfo {year}
  {2010})}\BibitemShut {NoStop}%
\bibitem [{\citenamefont {Barwood}\ \emph {et~al.}(2014)\citenamefont
  {Barwood}, \citenamefont {Huang}, \citenamefont {Klein}, \citenamefont
  {Johnson}, \citenamefont {King}, \citenamefont {Margolis}, \citenamefont
  {Szymaniec},\ and\ \citenamefont {Gill}}]{Barwood2014PRA}%
  \BibitemOpen
  \bibfield  {author} {\bibinfo {author} {\bibfnamefont {G.~P.}\ \bibnamefont
  {Barwood}}, \bibinfo {author} {\bibfnamefont {G.}~\bibnamefont {Huang}},
  \bibinfo {author} {\bibfnamefont {H.~A.}\ \bibnamefont {Klein}}, \bibinfo
  {author} {\bibfnamefont {L.~A.~M.}\ \bibnamefont {Johnson}}, \bibinfo
  {author} {\bibfnamefont {S.~A.}\ \bibnamefont {King}}, \bibinfo {author}
  {\bibfnamefont {H.~S.}\ \bibnamefont {Margolis}}, \bibinfo {author}
  {\bibfnamefont {K.}~\bibnamefont {Szymaniec}}, \ and\ \bibinfo {author}
  {\bibfnamefont {P.}~\bibnamefont {Gill}},\ }\href@noop {} {\bibfield
  {journal} {\bibinfo  {journal} {Phys. Rev. A}\ }\textbf {\bibinfo {volume}
  {89}},\ \bibinfo {pages} {050501} (\bibinfo {year} {2014})}\BibitemShut
  {NoStop}%
\bibitem [{\citenamefont {Huntemann}\ \emph {et~al.}(2016)\citenamefont
  {Huntemann}, \citenamefont {Sanner}, \citenamefont {Lipphardt}, \citenamefont
  {Tamm},\ and\ \citenamefont {Peik}}]{Huntemann2016PRL}%
  \BibitemOpen
  \bibfield  {author} {\bibinfo {author} {\bibfnamefont {N.}~\bibnamefont
  {Huntemann}}, \bibinfo {author} {\bibfnamefont {C.}~\bibnamefont {Sanner}},
  \bibinfo {author} {\bibfnamefont {B.}~\bibnamefont {Lipphardt}}, \bibinfo
  {author} {\bibfnamefont {C.}~\bibnamefont {Tamm}}, \ and\ \bibinfo {author}
  {\bibfnamefont {E.}~\bibnamefont {Peik}},\ }\href@noop {} {\bibfield
  {journal} {\bibinfo  {journal} {Phys. Rev. Lett.}\ }\textbf {\bibinfo
  {volume} {116}},\ \bibinfo {pages} {063001} (\bibinfo {year}
  {2016})}\BibitemShut {NoStop}%
\bibitem [{\citenamefont {Schmidt}\ \emph {et~al.}(2005)\citenamefont
  {Schmidt}, \citenamefont {Rosenband}, \citenamefont {Langer}, \citenamefont
  {Itano}, \citenamefont {Bergquist},\ and\ \citenamefont
  {Wineland}}]{Schmidt2005Science}%
  \BibitemOpen
  \bibfield  {author} {\bibinfo {author} {\bibfnamefont {P.~O.}\ \bibnamefont
  {Schmidt}}, \bibinfo {author} {\bibfnamefont {T.}~\bibnamefont {Rosenband}},
  \bibinfo {author} {\bibfnamefont {C.}~\bibnamefont {Langer}}, \bibinfo
  {author} {\bibfnamefont {W.~M.}\ \bibnamefont {Itano}}, \bibinfo {author}
  {\bibfnamefont {J.~C.}\ \bibnamefont {Bergquist}}, \ and\ \bibinfo {author}
  {\bibfnamefont {D.~J.}\ \bibnamefont {Wineland}},\ }\href@noop {} {\bibfield
  {journal} {\bibinfo  {journal} {Science}\ }\textbf {\bibinfo {volume}
  {309}},\ \bibinfo {pages} {749} (\bibinfo {year} {2005})}\BibitemShut
  {NoStop}%
\bibitem [{\citenamefont {Diedrich}\ \emph {et~al.}(1989)\citenamefont
  {Diedrich}, \citenamefont {Bergquist}, \citenamefont {Itano},\ and\
  \citenamefont {Wineland}}]{Diedrich1989PRL}%
  \BibitemOpen
  \bibfield  {author} {\bibinfo {author} {\bibfnamefont {F.}~\bibnamefont
  {Diedrich}}, \bibinfo {author} {\bibfnamefont {J.~C.}\ \bibnamefont
  {Bergquist}}, \bibinfo {author} {\bibfnamefont {W.~M.}\ \bibnamefont
  {Itano}}, \ and\ \bibinfo {author} {\bibfnamefont {D.~J.}\ \bibnamefont
  {Wineland}},\ }\href@noop {} {\bibfield  {journal} {\bibinfo  {journal}
  {Phys. Rev. Lett.}\ }\textbf {\bibinfo {volume} {62}},\ \bibinfo {pages}
  {403} (\bibinfo {year} {1989})}\BibitemShut {NoStop}%
\bibitem [{\citenamefont {Monroe}\ \emph {et~al.}(1995)\citenamefont {Monroe},
  \citenamefont {Meekhof}, \citenamefont {King}, \citenamefont {Jefferts},
  \citenamefont {Itano}, \citenamefont {Wineland},\ and\ \citenamefont
  {Gould}}]{Monroe1995PRL}%
  \BibitemOpen
  \bibfield  {author} {\bibinfo {author} {\bibfnamefont {C.}~\bibnamefont
  {Monroe}}, \bibinfo {author} {\bibfnamefont {D.~M.}\ \bibnamefont {Meekhof}},
  \bibinfo {author} {\bibfnamefont {B.~E.}\ \bibnamefont {King}}, \bibinfo
  {author} {\bibfnamefont {S.~R.}\ \bibnamefont {Jefferts}}, \bibinfo {author}
  {\bibfnamefont {W.~M.}\ \bibnamefont {Itano}}, \bibinfo {author}
  {\bibfnamefont {D.~J.}\ \bibnamefont {Wineland}}, \ and\ \bibinfo {author}
  {\bibfnamefont {P.}~\bibnamefont {Gould}},\ }\href@noop {} {\bibfield
  {journal} {\bibinfo  {journal} {Phys. Rev. Lett.}\ }\textbf {\bibinfo
  {volume} {75}},\ \bibinfo {pages} {4011} (\bibinfo {year}
  {1995})}\BibitemShut {NoStop}%
\bibitem [{\citenamefont {Roos}\ \emph {et~al.}(2000)\citenamefont {Roos},
  \citenamefont {Leibfried}, \citenamefont {Mundt}, \citenamefont
  {Schmidt-Kaler}, \citenamefont {Eschner},\ and\ \citenamefont
  {Blatt}}]{Roos2000PRL}%
  \BibitemOpen
  \bibfield  {author} {\bibinfo {author} {\bibfnamefont {C.~F.}\ \bibnamefont
  {Roos}}, \bibinfo {author} {\bibfnamefont {D.}~\bibnamefont {Leibfried}},
  \bibinfo {author} {\bibfnamefont {A.}~\bibnamefont {Mundt}}, \bibinfo
  {author} {\bibfnamefont {F.}~\bibnamefont {Schmidt-Kaler}}, \bibinfo {author}
  {\bibfnamefont {J.}~\bibnamefont {Eschner}}, \ and\ \bibinfo {author}
  {\bibfnamefont {R.}~\bibnamefont {Blatt}},\ }\href@noop {} {\bibfield
  {journal} {\bibinfo  {journal} {Phys. Rev. Lett.}\ }\textbf {\bibinfo
  {volume} {85}},\ \bibinfo {pages} {5547} (\bibinfo {year}
  {2000})}\BibitemShut {NoStop}%
\bibitem [{\citenamefont {Manfredi}\ and\ \citenamefont
  {Hervieux}(2012)}]{Manfredi2012PRL}%
  \BibitemOpen
  \bibfield  {author} {\bibinfo {author} {\bibfnamefont {G.}~\bibnamefont
  {Manfredi}}\ and\ \bibinfo {author} {\bibfnamefont {P.-A.}\ \bibnamefont
  {Hervieux}},\ }\href@noop {} {\bibfield  {journal} {\bibinfo  {journal}
  {Phys. Rev. Lett.}\ }\textbf {\bibinfo {volume} {109}},\ \bibinfo {pages}
  {255005} (\bibinfo {year} {2012})}\BibitemShut {NoStop}%
\bibitem [{\citenamefont {Lin}\ \emph {et~al.}(2013)\citenamefont {Lin},
  \citenamefont {Gaebler}, \citenamefont {Tan}, \citenamefont {Bowler},
  \citenamefont {Jost}, \citenamefont {Leibfried},\ and\ \citenamefont
  {Wineland}}]{Lin2013PRL}%
  \BibitemOpen
  \bibfield  {author} {\bibinfo {author} {\bibfnamefont {Y.}~\bibnamefont
  {Lin}}, \bibinfo {author} {\bibfnamefont {J.~P.}\ \bibnamefont {Gaebler}},
  \bibinfo {author} {\bibfnamefont {T.~R.}\ \bibnamefont {Tan}}, \bibinfo
  {author} {\bibfnamefont {R.}~\bibnamefont {Bowler}}, \bibinfo {author}
  {\bibfnamefont {J.~D.}\ \bibnamefont {Jost}}, \bibinfo {author}
  {\bibfnamefont {D.}~\bibnamefont {Leibfried}}, \ and\ \bibinfo {author}
  {\bibfnamefont {D.~J.}\ \bibnamefont {Wineland}},\ }\href@noop {} {\bibfield
  {journal} {\bibinfo  {journal} {Phys. Rev. Lett.}\ }\textbf {\bibinfo
  {volume} {110}},\ \bibinfo {pages} {153002} (\bibinfo {year}
  {2013})}\BibitemShut {NoStop}%
\bibitem [{\citenamefont {Ejtemaee}\ and\ \citenamefont
  {Haljan}(2016)}]{Ejtemaee2016SisyphusCooling}%
  \BibitemOpen
  \bibfield  {author} {\bibinfo {author} {\bibfnamefont {S.}~\bibnamefont
  {Ejtemaee}}\ and\ \bibinfo {author} {\bibfnamefont {P.~C.}\ \bibnamefont
  {Haljan}},\ }\href@noop {} {\  (\bibinfo {year} {2016})},\ \Eprint
  {http://arxiv.org/abs/1603.01248} {arXiv:1603.01248 [physics.atom-ph]}
  \BibitemShut {NoStop}%
\bibitem [{\citenamefont {Nisbet-Jones}\ \emph {et~al.}(2016)\citenamefont
  {Nisbet-Jones}, \citenamefont {King}, \citenamefont {Jones}, \citenamefont
  {Godun}, \citenamefont {Baynham}, \citenamefont {Bongs}, \citenamefont {M.},
  \citenamefont {Balling},\ and\ \citenamefont {Gill}}]{NisbetJones2016APB}%
  \BibitemOpen
  \bibfield  {author} {\bibinfo {author} {\bibfnamefont {P.~B.~R.}\
  \bibnamefont {Nisbet-Jones}}, \bibinfo {author} {\bibfnamefont {S.~A.}\
  \bibnamefont {King}}, \bibinfo {author} {\bibfnamefont {J.~M.}\ \bibnamefont
  {Jones}}, \bibinfo {author} {\bibfnamefont {R.~M.}\ \bibnamefont {Godun}},
  \bibinfo {author} {\bibfnamefont {C.~F.~A.}\ \bibnamefont {Baynham}},
  \bibinfo {author} {\bibfnamefont {K.}~\bibnamefont {Bongs}}, \bibinfo
  {author} {\bibfnamefont {D.}~\bibnamefont {M.}}, \bibinfo {author}
  {\bibfnamefont {P.}~\bibnamefont {Balling}}, \ and\ \bibinfo {author}
  {\bibfnamefont {P.}~\bibnamefont {Gill}},\ }\href@noop {} {\bibfield
  {journal} {\bibinfo  {journal} {Appl. Phys. B}\ }\textbf {\bibinfo {volume}
  {122}},\ \bibinfo {pages} {57} (\bibinfo {year} {2016})}\BibitemShut
  {NoStop}%
\bibitem [{\citenamefont {Poulsen}\ \emph {et~al.}(2012)\citenamefont
  {Poulsen}, \citenamefont {Miroshnychenko},\ and\ \citenamefont
  {Drewsen}}]{Poulsen2012PRA}%
  \BibitemOpen
  \bibfield  {author} {\bibinfo {author} {\bibfnamefont {G.}~\bibnamefont
  {Poulsen}}, \bibinfo {author} {\bibfnamefont {Y.}~\bibnamefont
  {Miroshnychenko}}, \ and\ \bibinfo {author} {\bibfnamefont {M.}~\bibnamefont
  {Drewsen}},\ }\href@noop {} {\bibfield  {journal} {\bibinfo  {journal} {Phys.
  Rev. A}\ }\textbf {\bibinfo {volume} {86}},\ \bibinfo {pages} {051402(R)}
  (\bibinfo {year} {2012})}\BibitemShut {NoStop}%
\bibitem [{\citenamefont {Wan}\ \emph {et~al.}(2015)\citenamefont {Wan},
  \citenamefont {Gebert}, \citenamefont {Wolf},\ and\ \citenamefont
  {Schmidt}}]{Wan2015PRA}%
  \BibitemOpen
  \bibfield  {author} {\bibinfo {author} {\bibfnamefont {Y.}~\bibnamefont
  {Wan}}, \bibinfo {author} {\bibfnamefont {F.}~\bibnamefont {Gebert}},
  \bibinfo {author} {\bibfnamefont {F.}~\bibnamefont {Wolf}}, \ and\ \bibinfo
  {author} {\bibfnamefont {P.~O.}\ \bibnamefont {Schmidt}},\ }\href@noop {}
  {\bibfield  {journal} {\bibinfo  {journal} {Phys. Rev. A}\ }\textbf {\bibinfo
  {volume} {91}},\ \bibinfo {pages} {043425} (\bibinfo {year}
  {2015})}\BibitemShut {NoStop}%
\bibitem [{SM()}]{SM}%
  \BibitemOpen
  \href@noop {} {}\bibinfo {howpublished} {Supplemental material.}\BibitemShut
  {Stop}%
\bibitem [{\citenamefont {Colombe}\ \emph {et~al.}(2014)\citenamefont
  {Colombe}, \citenamefont {Slichter}, \citenamefont {Wilson}, \citenamefont
  {Leibfried},\ and\ \citenamefont {Wineland}}]{Clolombe2014OE}%
  \BibitemOpen
  \bibfield  {author} {\bibinfo {author} {\bibfnamefont {Y.}~\bibnamefont
  {Colombe}}, \bibinfo {author} {\bibfnamefont {D.~H.}\ \bibnamefont
  {Slichter}}, \bibinfo {author} {\bibfnamefont {A.~C.}\ \bibnamefont
  {Wilson}}, \bibinfo {author} {\bibfnamefont {D.}~\bibnamefont {Leibfried}}, \
  and\ \bibinfo {author} {\bibfnamefont {D.~J.}\ \bibnamefont {Wineland}},\
  }\href@noop {} {\bibfield  {journal} {\bibinfo  {journal} {Opt. Express}\
  }\textbf {\bibinfo {volume} {22}},\ \bibinfo {pages} {19783} (\bibinfo {year}
  {2014})}\BibitemShut {NoStop}%
\bibitem [{\citenamefont {Barrett}\ \emph {et~al.}(2003)\citenamefont
  {Barrett}, \citenamefont {DeMarco}, \citenamefont {Schaetz}, \citenamefont
  {Meyer}, \citenamefont {Leibfried}, \citenamefont {Britton}, \citenamefont
  {Chiaverini}, \citenamefont {Itano}, \citenamefont {Jelenkovi\'c},
  \citenamefont {Jost}, \citenamefont {Langer}, \citenamefont {Rosenband},\
  and\ \citenamefont {Wineland}}]{Barrett2003PRA}%
  \BibitemOpen
  \bibfield  {author} {\bibinfo {author} {\bibfnamefont {M.~D.}\ \bibnamefont
  {Barrett}}, \bibinfo {author} {\bibfnamefont {B.}~\bibnamefont {DeMarco}},
  \bibinfo {author} {\bibfnamefont {T.}~\bibnamefont {Schaetz}}, \bibinfo
  {author} {\bibfnamefont {V.}~\bibnamefont {Meyer}}, \bibinfo {author}
  {\bibfnamefont {D.}~\bibnamefont {Leibfried}}, \bibinfo {author}
  {\bibfnamefont {J.}~\bibnamefont {Britton}}, \bibinfo {author} {\bibfnamefont
  {J.}~\bibnamefont {Chiaverini}}, \bibinfo {author} {\bibfnamefont {W.~M.}\
  \bibnamefont {Itano}}, \bibinfo {author} {\bibfnamefont {B.}~\bibnamefont
  {Jelenkovi\'c}}, \bibinfo {author} {\bibfnamefont {J.~D.}\ \bibnamefont
  {Jost}}, \bibinfo {author} {\bibfnamefont {C.}~\bibnamefont {Langer}},
  \bibinfo {author} {\bibfnamefont {T.}~\bibnamefont {Rosenband}}, \ and\
  \bibinfo {author} {\bibfnamefont {D.~J.}\ \bibnamefont {Wineland}},\
  }\href@noop {} {\bibfield  {journal} {\bibinfo  {journal} {Phys. Rev. A}\
  }\textbf {\bibinfo {volume} {68}},\ \bibinfo {pages} {042302} (\bibinfo
  {year} {2003})}\BibitemShut {NoStop}%
\bibitem [{\citenamefont {Wineland}\ \emph {et~al.}(1998)\citenamefont
  {Wineland}, \citenamefont {Monroe}, \citenamefont {Itano}, \citenamefont
  {Leibfried}, \citenamefont {King},\ and\ \citenamefont
  {Meekhof}}]{WinelandBible}%
  \BibitemOpen
  \bibfield  {author} {\bibinfo {author} {\bibfnamefont {D.~J.}\ \bibnamefont
  {Wineland}}, \bibinfo {author} {\bibfnamefont {C.}~\bibnamefont {Monroe}},
  \bibinfo {author} {\bibfnamefont {W.~M.}\ \bibnamefont {Itano}}, \bibinfo
  {author} {\bibfnamefont {D.}~\bibnamefont {Leibfried}}, \bibinfo {author}
  {\bibfnamefont {B.~E.}\ \bibnamefont {King}}, \ and\ \bibinfo {author}
  {\bibfnamefont {D.~M.}\ \bibnamefont {Meekhof}},\ }\href@noop {} {\bibfield
  {journal} {\bibinfo  {journal} {J. Res. Natl. Inst. Stand. Technol.}\
  }\textbf {\bibinfo {volume} {103}},\ \bibinfo {pages} {259} (\bibinfo {year}
  {1998})}\BibitemShut {NoStop}%
\bibitem [{\citenamefont {Turchette}\ \emph {et~al.}(2000)\citenamefont
  {Turchette}, \citenamefont {Myatt}, \citenamefont {King}, \citenamefont
  {Sackett}, \citenamefont {Kielpinski}, \citenamefont {Itano}, \citenamefont
  {Monroe},\ and\ \citenamefont {Wineland}}]{Turchette2000PRA}%
  \BibitemOpen
  \bibfield  {author} {\bibinfo {author} {\bibfnamefont {Q.~A.}\ \bibnamefont
  {Turchette}}, \bibinfo {author} {\bibfnamefont {C.~J.}\ \bibnamefont
  {Myatt}}, \bibinfo {author} {\bibfnamefont {B.~E.}\ \bibnamefont {King}},
  \bibinfo {author} {\bibfnamefont {C.~A.}\ \bibnamefont {Sackett}}, \bibinfo
  {author} {\bibfnamefont {D.}~\bibnamefont {Kielpinski}}, \bibinfo {author}
  {\bibfnamefont {W.~M.}\ \bibnamefont {Itano}}, \bibinfo {author}
  {\bibfnamefont {C.}~\bibnamefont {Monroe}}, \ and\ \bibinfo {author}
  {\bibfnamefont {D.~J.}\ \bibnamefont {Wineland}},\ }\href@noop {} {\bibfield
  {journal} {\bibinfo  {journal} {Phys. Rev. A}\ }\textbf {\bibinfo {volume}
  {62}},\ \bibinfo {pages} {053807} (\bibinfo {year} {2000})}\BibitemShut
  {NoStop}%
\bibitem [{\citenamefont {Ozeri}\ \emph {et~al.}(2007)\citenamefont {Ozeri},
  \citenamefont {Itano}, \citenamefont {Blakestad}, \citenamefont {Britton},
  \citenamefont {Chiaverini}, \citenamefont {Jost}, \citenamefont {Langer},
  \citenamefont {Leibfried}, \citenamefont {Reichle}, \citenamefont {Seidelin},
  \citenamefont {Wesenberg},\ and\ \citenamefont {Wineland}}]{Ozeri2007PRA}%
  \BibitemOpen
  \bibfield  {author} {\bibinfo {author} {\bibfnamefont {R.}~\bibnamefont
  {Ozeri}}, \bibinfo {author} {\bibfnamefont {W.~M.}\ \bibnamefont {Itano}},
  \bibinfo {author} {\bibfnamefont {R.~B.}\ \bibnamefont {Blakestad}}, \bibinfo
  {author} {\bibfnamefont {J.}~\bibnamefont {Britton}}, \bibinfo {author}
  {\bibfnamefont {J.}~\bibnamefont {Chiaverini}}, \bibinfo {author}
  {\bibfnamefont {J.~D.}\ \bibnamefont {Jost}}, \bibinfo {author}
  {\bibfnamefont {C.}~\bibnamefont {Langer}}, \bibinfo {author} {\bibfnamefont
  {D.}~\bibnamefont {Leibfried}}, \bibinfo {author} {\bibfnamefont
  {R.}~\bibnamefont {Reichle}}, \bibinfo {author} {\bibfnamefont
  {S.}~\bibnamefont {Seidelin}}, \bibinfo {author} {\bibfnamefont {J.~H.}\
  \bibnamefont {Wesenberg}}, \ and\ \bibinfo {author} {\bibfnamefont {D.~J.}\
  \bibnamefont {Wineland}},\ }\href@noop {} {\bibfield  {journal} {\bibinfo
  {journal} {Phys. Rev. A}\ }\textbf {\bibinfo {volume} {75}},\ \bibinfo
  {pages} {042329} (\bibinfo {year} {2007})}\BibitemShut {NoStop}%
\bibitem [{\citenamefont {Itano}\ and\ \citenamefont
  {Wineland}(1982)}]{Wayne1982PRA}%
  \BibitemOpen
  \bibfield  {author} {\bibinfo {author} {\bibfnamefont {W.~M.}\ \bibnamefont
  {Itano}}\ and\ \bibinfo {author} {\bibfnamefont {D.~J.}\ \bibnamefont
  {Wineland}},\ }\href@noop {} {\bibfield  {journal} {\bibinfo  {journal}
  {Phys. Rev. A}\ }\textbf {\bibinfo {volume} {25}},\ \bibinfo {pages} {35}
  (\bibinfo {year} {1982})}\BibitemShut {NoStop}%
\bibitem [{\citenamefont {Chou}\ \emph {et~al.}(2011)\citenamefont {Chou},
  \citenamefont {Hume}, \citenamefont {Thorpe}, \citenamefont {Wineland},\ and\
  \citenamefont {Rosenband}}]{Chou2011QCoherence}%
  \BibitemOpen
  \bibfield  {author} {\bibinfo {author} {\bibfnamefont {C.~W.}\ \bibnamefont
  {Chou}}, \bibinfo {author} {\bibfnamefont {D.~B.}\ \bibnamefont {Hume}},
  \bibinfo {author} {\bibfnamefont {M.~J.}\ \bibnamefont {Thorpe}}, \bibinfo
  {author} {\bibfnamefont {D.~J.}\ \bibnamefont {Wineland}}, \ and\ \bibinfo
  {author} {\bibfnamefont {T.}~\bibnamefont {Rosenband}},\ }\href@noop {}
  {\bibfield  {journal} {\bibinfo  {journal} {Phys. Rev. Lett.}\ }\textbf
  {\bibinfo {volume} {106}},\ \bibinfo {pages} {160801} (\bibinfo {year}
  {2011})}\BibitemShut {NoStop}%
\bibitem [{\citenamefont {Kessler}\ \emph {et~al.}(2012)\citenamefont
  {Kessler}, \citenamefont {Hagemann}, \citenamefont {Grebing}, \citenamefont
  {Legero}, \citenamefont {Sterr}, \citenamefont {Riehle}, \citenamefont
  {Martin}, \citenamefont {Chen},\ and\ \citenamefont {Ye}}]{Kessler2012NP}%
  \BibitemOpen
  \bibfield  {author} {\bibinfo {author} {\bibfnamefont {T.}~\bibnamefont
  {Kessler}}, \bibinfo {author} {\bibfnamefont {C.}~\bibnamefont {Hagemann}},
  \bibinfo {author} {\bibfnamefont {C.}~\bibnamefont {Grebing}}, \bibinfo
  {author} {\bibfnamefont {T.}~\bibnamefont {Legero}}, \bibinfo {author}
  {\bibfnamefont {U.}~\bibnamefont {Sterr}}, \bibinfo {author} {\bibfnamefont
  {F.}~\bibnamefont {Riehle}}, \bibinfo {author} {\bibfnamefont {M.~J.}\
  \bibnamefont {Martin}}, \bibinfo {author} {\bibfnamefont {L.}~\bibnamefont
  {Chen}}, \ and\ \bibinfo {author} {\bibfnamefont {J.}~\bibnamefont {Ye}},\
  }\href@noop {} {\bibfield  {journal} {\bibinfo  {journal} {Nature Photon.}\
  }\textbf {\bibinfo {volume} {6}},\ \bibinfo {pages} {687} (\bibinfo {year}
  {2012})}\BibitemShut {NoStop}%
\bibitem [{\citenamefont {H\"afner}\ \emph {et~al.}(2015)\citenamefont
  {H\"afner}, \citenamefont {Falke}, \citenamefont {Grebing}, \citenamefont
  {Vogt}, \citenamefont {Legero},\ and\ \citenamefont
  {Merimaa}}]{Hafner2015OL}%
  \BibitemOpen
  \bibfield  {author} {\bibinfo {author} {\bibfnamefont {S.}~\bibnamefont
  {H\"afner}}, \bibinfo {author} {\bibfnamefont {S.}~\bibnamefont {Falke}},
  \bibinfo {author} {\bibfnamefont {C.}~\bibnamefont {Grebing}}, \bibinfo
  {author} {\bibfnamefont {S.}~\bibnamefont {Vogt}}, \bibinfo {author}
  {\bibfnamefont {T.}~\bibnamefont {Legero}}, \ and\ \bibinfo {author}
  {\bibfnamefont {M.}~\bibnamefont {Merimaa}},\ }\href@noop {} {\bibfield
  {journal} {\bibinfo  {journal} {Opt. Lett.}\ }\textbf {\bibinfo {volume}
  {40}},\ \bibinfo {pages} {2112} (\bibinfo {year} {2015})}\BibitemShut
  {NoStop}%
\bibitem [{\citenamefont {Hume}\ and\ \citenamefont
  {Leibrandt}(2016)}]{Hume2016PRA}%
  \BibitemOpen
  \bibfield  {author} {\bibinfo {author} {\bibfnamefont {D.~B.}\ \bibnamefont
  {Hume}}\ and\ \bibinfo {author} {\bibfnamefont {D.~R.}\ \bibnamefont
  {Leibrandt}},\ }\href@noop {} {\bibfield  {journal} {\bibinfo  {journal}
  {Phys. Rev. A}\ }\textbf {\bibinfo {volume} {93}},\ \bibinfo {pages} {032138}
  (\bibinfo {year} {2016})}\BibitemShut {NoStop}%
\bibitem [{\citenamefont {Gaebler}\ \emph {et~al.}(2016)\citenamefont
  {Gaebler}, \citenamefont {Tan}, \citenamefont {Lin}, \citenamefont {Wan},
  \citenamefont {Bowler}, \citenamefont {Keith}, \citenamefont {Glancy},
  \citenamefont {Coakley}, \citenamefont {Knill}, \citenamefont {Leibfried},\
  and\ \citenamefont {Wineland}}]{Gaebler2016HFgate}%
  \BibitemOpen
  \bibfield  {author} {\bibinfo {author} {\bibfnamefont {J.~P.}\ \bibnamefont
  {Gaebler}}, \bibinfo {author} {\bibfnamefont {T.~R.}\ \bibnamefont {Tan}},
  \bibinfo {author} {\bibfnamefont {Y.}~\bibnamefont {Lin}}, \bibinfo {author}
  {\bibfnamefont {Y.}~\bibnamefont {Wan}}, \bibinfo {author} {\bibfnamefont
  {R.}~\bibnamefont {Bowler}}, \bibinfo {author} {\bibfnamefont {A.~C.}\
  \bibnamefont {Keith}}, \bibinfo {author} {\bibfnamefont {S.}~\bibnamefont
  {Glancy}}, \bibinfo {author} {\bibfnamefont {K.}~\bibnamefont {Coakley}},
  \bibinfo {author} {\bibfnamefont {E.}~\bibnamefont {Knill}}, \bibinfo
  {author} {\bibfnamefont {D.}~\bibnamefont {Leibfried}}, \ and\ \bibinfo
  {author} {\bibfnamefont {D.~J.}\ \bibnamefont {Wineland}},\ }\href@noop {}
  {\bibfield  {journal} {\bibinfo  {journal} {Phys. Rev. Lett}\ }\textbf
  {\bibinfo {volume} {117}},\ \bibinfo {pages} {060505} (\bibinfo {year}
  {2016})}\BibitemShut {NoStop}%
\bibitem [{\citenamefont {DiCiccio}\ and\ \citenamefont
  {Efron}(1996)}]{Diciccio1996SS}%
  \BibitemOpen
  \bibfield  {author} {\bibinfo {author} {\bibfnamefont {T.~J.}\ \bibnamefont
  {DiCiccio}}\ and\ \bibinfo {author} {\bibfnamefont {B.}~\bibnamefont
  {Efron}},\ }\href@noop {} {\bibfield  {journal} {\bibinfo  {journal}
  {Statist. Sci.}\ }\textbf {\bibinfo {volume} {11}},\ \bibinfo {pages} {189}
  (\bibinfo {year} {1996})}\BibitemShut {NoStop}%
\bibitem [{\citenamefont {James}(1998)}]{James1998PRL}%
  \BibitemOpen
  \bibfield  {author} {\bibinfo {author} {\bibfnamefont {D.~F.}\ \bibnamefont
  {James}},\ }\href@noop {} {\bibfield  {journal} {\bibinfo  {journal} {Phys.
  Rev. Lett.}\ }\textbf {\bibinfo {volume} {81}},\ \bibinfo {pages} {317}
  (\bibinfo {year} {1998})}\BibitemShut {NoStop}%
\bibitem [{\citenamefont {Dodonov}\ \emph {et~al.}(2000)\citenamefont
  {Dodonov}, \citenamefont {Mizrahi},\ and\ \citenamefont
  {de~Souza~Silva}}]{Dodonov2000JOB}%
  \BibitemOpen
  \bibfield  {author} {\bibinfo {author} {\bibfnamefont {V.~V.}\ \bibnamefont
  {Dodonov}}, \bibinfo {author} {\bibfnamefont {S.~S.}\ \bibnamefont
  {Mizrahi}}, \ and\ \bibinfo {author} {\bibfnamefont {A.~L.}\ \bibnamefont
  {de~Souza~Silva}},\ }\href@noop {} {\bibfield  {journal} {\bibinfo  {journal}
  {J. Opt. B: Quantum Semiclass. Opt.}\ }\textbf {\bibinfo {volume} {2}},\
  \bibinfo {pages} {271} (\bibinfo {year} {2000})}\BibitemShut {NoStop}%
\bibitem [{\citenamefont {Wineland}\ \emph {et~al.}(1987)\citenamefont
  {Wineland}, \citenamefont {Itano}, \citenamefont {Bergquist},\ and\
  \citenamefont {Hulet}}]{Wineland1987PRA}%
  \BibitemOpen
  \bibfield  {author} {\bibinfo {author} {\bibfnamefont {D.~J.}\ \bibnamefont
  {Wineland}}, \bibinfo {author} {\bibfnamefont {W.}~\bibnamefont {Itano}},
  \bibinfo {author} {\bibfnamefont {J.}~\bibnamefont {Bergquist}}, \ and\
  \bibinfo {author} {\bibfnamefont {R.}~\bibnamefont {Hulet}},\ }\href@noop {}
  {\bibfield  {journal} {\bibinfo  {journal} {Phys. Rev. A}\ }\textbf {\bibinfo
  {volume} {36}},\ \bibinfo {pages} {2220} (\bibinfo {year}
  {1987})}\BibitemShut {NoStop}%
\end{thebibliography}
%

\end{document}